\documentclass[sigconf]{acmart}

\AtBeginDocument{%
  \providecommand\BibTeX{{%
    \normalfont B\kern-0.5em{\scshape i\kern-0.25em b}\kern-0.8em\TeX}}}

\usepackage{balance}
\usepackage{wrapfig}
\acmSubmissionID{8648}

\copyrightyear{2022} 
\acmYear{2022} 
\setcopyright{acmcopyright}\acmConference[CHI '22]{CHI Conference on Human Factors in Computing Systems}{April 29-May 5, 2022}{New Orleans, LA, USA}
\acmBooktitle{CHI Conference on Human Factors in Computing Systems (CHI '22), April 29-May 5, 2022, New Orleans, LA, USA}
\acmPrice{15.00}
\acmDOI{10.1145/3491102.3517712}
\acmISBN{978-1-4503-9157-3/22/04}
\begin{document}

\title{RoleSeer: Understanding Informal Social Role Changes in MMORPGs via Visual Analytics}


\author{Laixin Xie}
\affiliation{%
  \institution{School of Information Science and Technology, ShanghaiTech University}
  \city{Shanghai}
  \country{China}}
\email{xielx@shanghaitech.edu.cn}

\author{Ziming Wu}
\affiliation{%
  \institution{Interactive Entertainment Group, Tencent Inc.}
  \city{Shenzhen}
  \country{China}}
  \email{zwual@connect.ust.hk}

\author{Peng Xu}
\affiliation{%
 \institution{User Experience Center, NetEase Inc.}
 \city{Hangzhou}
 \country{China}}
 \email{alexkdd@163.com}

\author{Wei Li}
\affiliation{%
  \institution{College of Creative Design, Shenzhen Technology University}
  \city{Shenzhen}
  \country{China}}
  \email{helloweili@icloud.com}

\author{Xiaojuan Ma}
\affiliation{%
  \institution{Department of Computer Science and Engineering, The Hong Kong University of Science and Technology}
  \city{Hong Kong}
  \country{China}}
\email{mxj@cse.ust.hk}

\author{Quan Li}
\authornote{The corresponding author.}
\affiliation{%
  \institution{School of Information Science and Technology, ShanghaiTech University}
  \city{Shanghai}
  \country{China}}
\email{liquan@shanghaitech.edu.cn}

\renewcommand{\shortauthors}{Xie, L., Wu, Z., Xu, P., Li, W., Ma, X., Li, Q.}

\begin{abstract}
Massively multiplayer online role-playing games create virtual communities that support heterogeneous ``social roles'' determined by gameplay interaction behaviors under a specific social context. For all social roles, formal roles are pre-defined, obvious, and explicitly ascribed to the people holding the roles, whereas informal roles are not well-defined and unspoken. Identifying the informal roles and understanding their subtle changes are critical to designing sociability mechanisms. However, it is nontrivial to understand the existence and evolution of such roles due to their loosely defined, interconvertible, and dynamic characteristics. We propose a visual analytics system, \textit{RoleSeer}, to investigate informal roles from the perspectives of behavioral interactions and depict their dynamic interconversions and transitions. Two cases, experts' feedback, and a user study suggest that \textit{RoleSeer} helps interpret the identified informal roles and explore the patterns behind role changes. We see our approach's potential in investigating informal roles in a broader range of social games.
\end{abstract}



\maketitle

\section{Introduction}
\par Like in real-life communities, heterogeneous social dynamics are also observed in virtual communities. Massively multiplayer online role-playing games (MMORPGs) are essentially virtual communities in the form of role-playing. As the gameplay takes place, members of the community adopt different social roles (i.e., ``\textit{a set of expectations and norms that define how people playing the roles should and would behave in a social situation}''~\cite{ashforth2000role,herrmann2004role}). Some roles are public and official. They can be easily observed from the given positions or titles, often labeled as \textit{warriors}, \textit{mages}, \textit{hunters} in games like \textit{World of Warcraft (WoW)}. As these roles are evidently pronounced by the game mechanism and independent from the individual behavior patterns, they are often recognized as \textit{formal roles}~\cite{ang2010social}. However, compared to the formal roles, some roles are less evident but can be equally important. For instance, a member's social status is primarily influenced by its contribution to the community. When an old member helps a newcomer, it adds to the effort of building a healthy community culture -- Altruism behaviors like such fortify the sense of belonging, which will eventually benefit the community as an entirety. Alternatively, players may also only take periphery positions and play as bystanders. The difference of these acts may marginally affect the formal roles but subtly shift the community's social structure. Based on players' respective influence upon the social attributes, roles such as \textit{dungeon leader}, \textit{fighter}, \textit{isolate}, \textit{social butterflies}, and \textit{information giver} are naturally assigned based on their explicit play behaviors~\cite{ang2010social,ashforth2000role,golder2004social,herrmann2004role}. Following social studies~\cite{ashforth2000role,herrmann2004role}, we use \textit{informal roles}\footnote{https://eagle.northwestu.edu/faculty/gary-gillespie/roles-and-small-group-communication/} to represent the later, less evident roles, which are naturally determined by players' gameplay behaviors.

\par Characterizing the collective configuration of informal roles in a community and understanding the existence and changes of roles within are essential ways to study MMORPG communities. First, informal roles are greatly influenced by existing formal roles. They will often replace or supplement formal social roles~\cite{smith1966observations}, which can provide insights into the sociability design of gameplay mechanisms that can facilitate the formation of specific informal roles~\cite{anderson2004dimensions}. Second, analyzing how individuals' informal roles vary regarding their behaviors can promote the long-term operations of a virtual community, such as drawing in lapsed gamers and encouraging, e.g., ``isolate'' players to change their roles.

\par However, identifying informal roles and understanding the reasons behind their existence and changes is nontrivial due to several issues: \textbf{(1) Loose definition.} Existing literature has attempted to provide descriptions of informal roles in MMORPGs. For example, an early study on text-based multi-user dungeons (often referred to as the predecessor of MMORPGs) categorized the typology of roles into four groups~\cite{bartle1996hearts}: \textit{killers} who like to annoy other players; \textit{socializers} who want causal social interaction with other players; \textit{achievers} who aim to master the game, and \textit{explorers} who enjoy exploring the game world. Although they provide grounded theories for later research on social roles in the context of graphical MMORPGs, the current definitions of the limited set of informal roles are primarily based on prior knowledge of game designers and analysts. Consequently, existing methods for deriving meaningful roles or patterns using pure statistical analysis approaches are insufficient without a delineated definition of informal roles, especially for those unspoken ones, making automated solutions of informal roles mining challenging to achieve. \textbf{(2) Dynamic interconversion and evolution.} Informal roles are based on human needs for trust, support, resource sharing irrespective of whether they operate in real life or a virtual setting~\cite{lai2008groups}. Specifically, these roles result from social interaction and negotiation between the actor and those they interact with. They are dynamically developing and interconvertible over time, i.e., players may switch from one role to another. Given such characteristics of informal roles, summarizing and understanding the temporal properties of the dynamic interconversion and evolution of informal roles in an MMORPG virtual community can help reveal players' exploration patterns and shed light upon the effective socialization mechanism of the players from a global perspective. \textbf{(3) Diverse paths behind role formations.} Diverse steps of social exploration (i.e., social self-discovery helping an individual understand their own social needs~\cite{jenkins2000categorization}) precede forming a particular informal role. For example, players may need to balance their personal and social time or choose between large but weak social ties and a few strong social ties when interacting with other virtual companions in the game~\cite{moreland1982socialization,morrison2002newcomers}. Although a small number of players with abundant external resources (e.g., time and money) could take on specific informal roles through the spending of these resources, the large majority of the members in the game community may not have sufficient external resources at their disposal. Instead, players adopt some informal roles through gameplay interactions. Inspecting the diverse paths behind the formations of such informal roles is more important because they can explain how people interact, collaborate, and work together to cultivate community building and growth. It can also increase the participant's awareness of their social interaction~\cite{strijbos2004effect}.

\par Most previous work tends to study the social role from ``who the users are'' and then ``what the users do'' rather than behavioral interactions between them in the virtual community, or they assumed a prior knowledge about the social roles (e.g., focusing on the contents of the players' posting or activities)~\cite{ang2010social,canossa2019influencers,williams2014structural}. Although such studies have been helpful, we argue that focusing on behavioral interactions would cast new light on social roles in virtual worlds. Thus, to fill the gap, we propose \textit{RoleSeer}, an interactive visual analytics system that helps game designers and game user experience (UX) practitioners understand informal roles and paths behind their formations and dynamic changes in the context of an MMORPG. To give detailed elaborations on the proposed idea and follow-up implementations, we organize this work as follows. We first observe our collaborating game team's current gaming social role analysis practices and identify their primary needs and concerns. Then, we adapt a dynamic network embedding and alignment approach to the friendship network in the specified MMORPG for facilitating potential informal role detection across multiple time snapshots. Different clusters of potential informal roles and their interconversion and evolution across these clusters are discovered by projecting the resulting embeddings onto a low-dimensional space. We further support the experts to explore players' diverse behavioral interactions that lead to their role changes. Based on these objectives, we develop a visual analytics system to support fine-grained informal role analysis at the overview, role cluster, and individual levels. Lastly, we present several case studies and interview feedback with domain experts to evaluate the efficacy of our system. We outline the contributions of this work as follows.
\begin{itemize}
\item We shadow domain experts' daily working processes and conduct interviews to get insight into their current practice in understanding informal role changes. 
\item We identify the potential informal roles from the perspective of behavioral interaction analysis through an adapted dynamic network embedding and alignment model in the context of an MMORPG.
\item We depict the interconversion and evolution of informal roles across different time snapshots and explore the patterns behind the role changes via a visual analytics system. 
\end{itemize}

\section{Related Work}
\par Literature that overlaps this work can be categorized into three groups: \textit{social network analysis in MMORPGs}, \textit{graph latent representations}, and \textit{dynamic graph visualization}.

\subsection{Social Network Analysis in MMORPGs}
\par Studies that focus on social network analysis in MMORPGs suggest that players' social behaviors and interactions considerably influence players' gaming experience~\cite{ducheneaut2006alone,jia2015socializing,kirman2009gaming,wohn2010building}. For example, Szell and Thurner~\cite{szell2010measuring} studied the structure of friend, enemy, and communication networks and identified that friend and enemy networks are topologically different. Ducheneaut et al.~\cite{ducheneaut2007life} and Chen et al.~\cite{chen2008player} used traditional metrics in social network analysis such as \textit{density} and \textit{centrality} to analyze the properties of player guilds in \textit{WoW}. Lu et al.~\cite{lu2019visual} proposed \textit{BeXplorer} to explore the dynamic interplay among multiple types of behaviors. Li et al.~\cite{li2017visual} investigated the evolution of egocentric players and focused on the relationship between a player (ego) and his/her directly linked friends (alters). They also inferred how changes in an ego's interactive behaviors might propagate through the friendship network. However, their work captured the evolutionary pattern based on ego networks at a microscopic level, a case-by-case analysis. Our work identifies representative informal roles from the perspectives of structural analysis at a global level. Furthermore, we study the underlying patterns that drive the interconversion and transition of informal roles held by players across multiple time snapshots.

\par Apart from studying the overall social relationships within a virtual community, other studies attempted to investigate players' social status or social roles such as \textit{leader}, \textit{core members}, and \textit{newcomers}~\cite{ang2010social}. For example, Williams et al.~\cite{williams2014structural} conducted interviews with hardcore players and demonstrated the importance of leaders and critical structural positions within a virtual community. Canossa et al.~\cite{canossa2019influencers} applied standard social network features to identify ``influencers'' to an online multiplayer shooter game. The result shows that network feature-defined influencers had an outsized impact on the playtime and social play of players joining their in-game network. These studies shed light upon the social role analysis in the game community. However, the previous social roles are pre-defined in an initial set of categories such as ``influencer'' and ``hardcore player'' based on experts' prior knowledge or manually categorized using thematic and content analysis~\cite{ang2010social}, which cannot identify more unspoken informal roles. Ang et al.~\cite{ang2010social} observed the ``chat interaction'' to determine the structural characteristics of three social roles of a guild community in \textit{WoW}: densely connected core members, loosely connected semi-periphery members, and disconnected periphery players. However, they only focused on a small guild in one static snapshot and verbal interactions among players. In our work, we identify informal roles based on behavioral interactions. We study the players' non-verbal interaction and explain the interconversion and transitions of informal roles. Ducheneaut et al.~\cite{10.1145/1124772.1124834} analyzed the prevalence and extent of social activities to investigate whether (and how) a game's ``social factor'' can contribute to the game's success. They found that playing the game is like being ``alone together,'' i.e., surrounded by others but not necessarily actively interacting with them. Their study suggests designs for online games in which encouraging and supporting direct interactions might be ``less important'' than designing for the ``spectator experience'' and a sense of social presence, indicating different social interactions and roles may be equally important to the game's success. Motivated by their work, we study informal roles' existence and dynamic transitions for a better user experience.

\subsection{Graph Latent Representations}
\par Graph latent representations have demonstrated their practicality in many graph analysis applications and downstream machine learning tasks such as node classification, clustering, and link prediction~\cite{cai2018comprehensive}. Perozzi et al.~\cite{perozzi2014deepwalk} developed a random walk-based method, \textit{DeepWalk}, to learn latent representations for nodes by generalizing neural language models to preserve the higher-order proximity between nodes. \textit{Node2Vec}~\cite{grover2016node2vec} provides a trade-off between breadth-first search (BFS) and depth-first search (DFS) in the random walk process. Van den Elzen et al.~\cite{van2015reducing} derived snapshot representations from adjacency matrices and then projected the snapshots onto a 2D space for network state discovery. Dal Col et al.~\cite{dal2017wavelet} used the coefficients of graph wavelets to represent nodes for evolutionary local change discovery. Xu et al.~\cite{xu2018exploring} exploited diachronic node embeddings (DNE) for better preserving structural and temporal properties. They further designed a system for an interactive exploration of dynamic networks. Particularly, its backend model first extracts node embeddings of different timestamps and then applies embedding alignment, which shows strong capability in capturing local information, i.e., the similarity among players. Eren et al.~\cite{eren2020dg2pix} displayed the similarity and difference of temporally dynamic graphs by sorting the graph embedding vectors. Our approach is different from the above works on two dimensions. First, they applied \textit{DeepWalk} to the network at each time snapshot and initialized the embedding results to preserve the temporal community, while we replaced the underlying \textit{DeepWalk} with \textit{Struc2Vec}~\cite{ribeiro2017struc2vec} for baseline model comparison. The adapted version can maintain the structural similarity that captures the ``roles" of nodes in the graph. Second, we focus on the interaction patterns that explain informal roles' existence, interconversion, and transitions.

\subsection{Dynamic Graph Visualization}
\par Dynamic graph visualization has been attracting researchers' attention for a long time. Two primary visualization methods have been proposed: \textit{animation} and \textit{timeline-based methods}. Particularly, the former method simulates the graph dynamic evolution by redrawing the graph at each step, visualizing it with a node-link diagram~\cite{beck2017taxonomy}, showing its structural characteristics. However, animation techniques are desirable to reduce the complexity of dynamic graphs to facilitate visualization; however, they are inadequate to support detailed network analysis and interpretation of temporal properties~\cite{archambault2010animation}. Meanwhile, users must maintain a mental map in each snapshot for comparison~\cite{archambault2010animation,bach2013graphdiaries}. Instead of animating the network as a sequence, timeline-based techniques directly draw the network in a static image at each time step along a timeline, providing a better temporal overview of the dynamic graph~\cite{groh2009interactively}. Small multiples such as matrices~\cite{perer2012matrixflow} and node-link diagrams~\cite{beck2017taxonomy} are employed to study position changes of nodes within the dynamic graph~\cite{van2013dynamic}. Although these methods can capture temporal structural changes, human cognition often limits the discovery of temporal patterns. Furthermore, when analyzing an extensive network, one may encounter scalability difficulties when space is limited~\cite{federico2012visual,pohl2008time}. In this work, we combine timeline visualization with our graph latent representations. Precisely, the latent graph representation captures the structural similarity and dynamics within a short period (e.g., $6$ hours for one step). The timeline aligns the corresponding informal roles with visualizing the overall trend across the entire time steps. Thus, we could accurately capture the dynamics since we analyze the graph in a much shorter period while still maintaining an overview of the overall changes, achieving a balance of detail and abstraction.

\section{Observational Study}
\par This section presents the background information of the showcased game and the game team we collaborate with. Then, we shadow the experts' daily working process and summarize their conventional practice and bottlenecks. Finally, we distill the design requirements based on the results of expert interviews. 

\begin{table*}[h]
\caption{Event and intimacy: $X$ indicates varying intimacy values of different events. $N$ shows no upper bound.}
\vspace{-3mm}
\label{tab:intimacy}
\centering \small
\begin{tabular}{c|c|c|c|c|c|c|c|c}
\hline \hline
     & \textit{killing monster}      & \textit{killing player}    & \textit{task} & \textit{Using props} & \textit{fighting} & \textit{chatting} & \textit{carbon} & \textit{battle} \\ \hline \hline
each increase       & 1          &  1         & X          & X           & 25 & 2   & X & 120      \\
weekly bound &    1500       & 300          & 100          & 5000          & 350           & 210 & N & N        \\ \hline
\end{tabular}
\end{table*}

\subsection{About the Game and the Team}
\par The game studied in this work is a typical MMORPG that provides various systems to facilitate social interactions among players, such as text/voice chatting and in-game activities. Among all the systems, the friendship system is the most popular one, allowing players to finish specific tasks, e.g., adding others as friends, seeking help, or finding teammates for challenging dungeons. Typically, friendship closeness is measured by the ``intimacy value'' between two players, designed by the game team to evaluate the degree of social interaction between two players. Different interactions such as \textit{chatting}, \textit{killing monsters}, and \textit{battles} among players contribute to varying intimacy values (\autoref{tab:intimacy}). 

\begin{figure}[h]
 \centering 
 \includegraphics[width=0.5\textwidth]{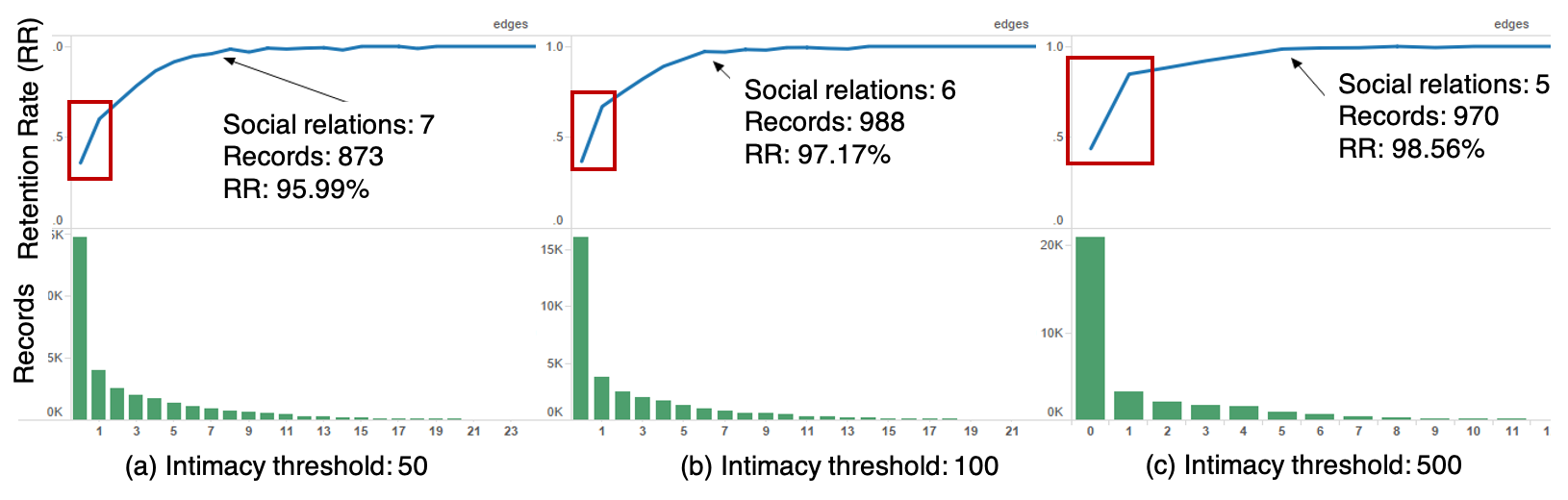}
  \vspace{-6mm}
 \caption{The impact of the number of players' social relations established in the first week on the players' retention rate in the next week: x-axis indicates the number of social ties found in the first week. The top half y-axis shows the retention rate of players in the next week, and the bottom half y-axis is the number of the involved players. Three red rectangles indicate the retention differences between players with $0$ and $1$ social relation in terms of three intimacy thresholds (i.e., different values of intimacy).}
 \label{fig:retention}
\end{figure}

\par We collaborated with a team of experts from an internet game company, including one user experience (UX) engineer (E1, female, age: $28$), one data analyst (E2, male, age: $29$), and two game designers (E3, male, age: $29$, E4, male, age: $30$), to study players' socialization in the MMORPG as mentioned above. All experts have been in the game industry for more than five years, and they have been involved in this specific MMORPG since its preliminary inception. One of the primary responsibilities of the game team is constantly improving the game's retention rate. According to a preliminary analysis of the game team (\autoref{fig:retention}), players with certain social relations have a 20\% higher retention rate than those without social ties (indicated by the three red rectangles). Meanwhile, specific turning points can be witnessed. For example, players with $5$ -- $7$ social relations have a considerably higher retention rate than those with fewer social ties. The experts commented that social ties could help improve players' retention, thus considering $5$ -- $7$ social relations the target value of players' socialization guidance in the early gaming stage. This initial observation also motivates our work. The game team wanted to observe how players establish social relations and get involved in the community, so they could design better social mechanisms to support players' socialization.

\subsection{Experts' Conventional Practice and Bottlenecks}

\par
\begin{figure}[h]
 \centering 
 \includegraphics[width=0.5\textwidth]{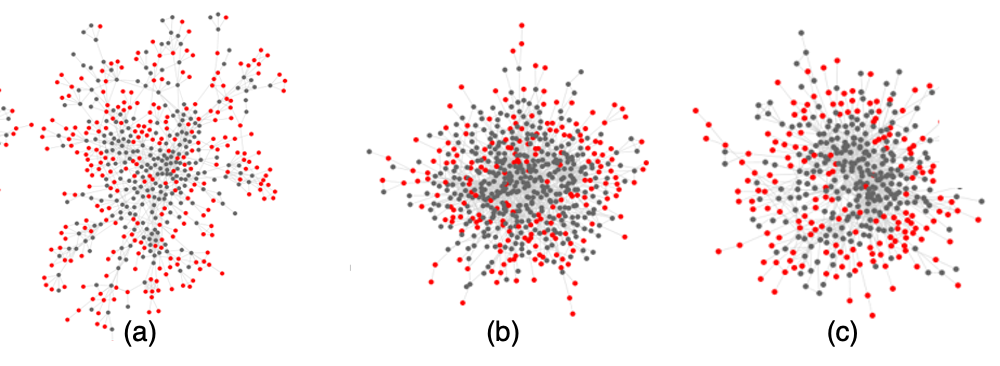}
  \vspace{-6mm}
 \caption{The force-directed layout of the social networks in three virtual communities. (a) PAs are well integrated into the community. (b -- c) PAs are mainly distributed in the peripheral areas and have fewer social relations.}
 \label{fig:community}
    \vspace{-3mm}
\end{figure}

 To obtain detailed information of the game experts' current practice, with consent, we shadowed the team's daily working process, including videotaping how they observed players experiencing the game, conducting testing experiments, and on-site interviews with the players. Later, we carried out a retrospective analysis with the game team on their conventional practices. Particularly, to tackle the issue mentioned above, the game team attempted to use a node-link diagram to visualize how players establish their social relations over time. They found that at the beginning of the game, three or four players may set up a small group, and the connections within the group become close. Then, the small group begins to establish social contact with external players, followed by integration with another group. Finally, the newly formed group integrates into a larger one. The experts identified that most players (approximately 60\%) follow a similar procedure in-game socialization, and their topological positions in the entire social network are quite different. For example, in \autoref{fig:community}, red dots indicate players entering the game after the first day of game release (marked as PAs), and black dots indicate players entering the game on the first day (marked as PBs). E1 compared the top three largest communities and drew the following three conclusions. First, in the first network (\autoref{fig:community} (a)), the proportion of PAs is higher compared with that in the other two networks (\autoref{fig:community} (bc)). Second, most PAs are in the central areas of the network (\autoref{fig:community} (a)). Third, PAs establish more social relations in \autoref{fig:community} (a), i.e., red dots with only one social tie are few. Many nodes have been integrated into groups or have established multiple social relations. In \autoref{fig:community} (bc), PAs are mainly distributed in the peripheral areas with few social relations.

\begin{figure*}[h]
 \centering 
 \includegraphics[width=\textwidth]{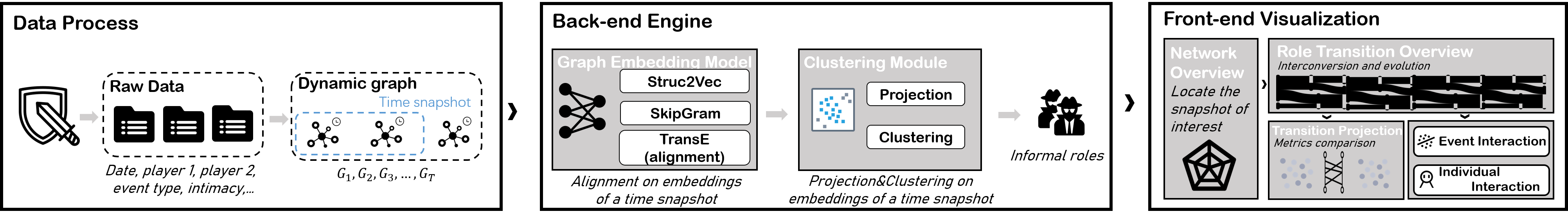}
  \vspace{-6mm}
 \caption{Pipeline of \textit{RoleSeer}: (1) data processing to produce dynamic graphs; (2) back-end engine consists of a graph embedding model and a clustering module to identify informal roles, and (3) front-end visualization facilitates explorative data analysis.}
 \label{fig:pipeline}
   \vspace{-3mm}
\end{figure*}

\par Although the experts obtained some initial insights into how players' socialization affects the integration into the community, they encountered the following issues when trying to make sense of players' exact social positions and the underlying reasons that lead to their social positions. First, the game team distinguished players based on ``who the players are'' and ``what the players do'' and compared their social positions from an overview perspective. The experts felt it insufficient to analyze players' structural similarities and differences. Second, from the obtained node-link visualizations, the experts could observe the changes of players' social positions, e.g., ``\textit{some players are moving from the peripheral positions to more central areas while others remain unchanged}'' (E1). Nevertheless, given the complexity and dynamics of the network, they have no idea how to track the movement of players' role changes in detail. Third, the current practice enabled the experts to observe the retention rates of different players (e.g., players with a different number of social relations). However, they had no idea what the collective interactions are that lead to such role differences, ``\textit{if we know what they are, we could design more appropriate social mechanisms,}'' said E3.

\subsection{Experts' Needs and Expectations}
\par We interviewed E1 -- E2 in one session and E3 -- E4 in another session to identify their primary concerns about analyzing players' social status and actions that contribute to social roles. Each session for the participants lasted about half an hour. At the end of the interviews, the need for a visual analytics system to ground the team's conversation with social role analysis emerged as a key theme. We summarize their requirements as follows.

\textbf{R1 Distinguish potential informal roles.} According to the experts' feedback, defining informal social roles largely depends on the experts' prior knowledge, assuming that some typical informal roles can be quickly formalized from the perspective of ``conventional'' social science. For instance, the experts would design rules based on what they did to identify the players. ``\textit{Although the designed rules can classify players holding different roles, the approach can be expensive and cumbersome, and uncertainty exists whether more effective rules could be extracted to cover more social roles,}'' said E2. Therefore, they wished for a more effective way to distinguish potential informal roles. 

\textbf{R2 Demonstrate information of corresponding informal roles.} After potential informal roles are identified, a subsequent analytical task raised by the experts is to interpret these roles. For instance, E1 pointed out that ``\textit{we need to determine how the role members are positioned and interact with their neighbors.}'' E2 commented that ``\textit{we should be able to observe the similarities and differences among the detected informal roles.}'' Specifically, each informal role should be inspected to allow the experts to take a closer look at their local social structure information and characteristics. 

\textbf{R3 Display interconversion and transition of informal roles.} Uncovering dynamic interconversion and transition among informal roles can facilitate summarizing the temporal patterns in the virtual community within a particular period. For example, E1 wanted to identify the evolving trajectories among the informal roles, which can demonstrate the role changes of players and the accompanying statistical information.

\textbf{R4 Track the evolution of informal roles over time.} All experts mentioned that investigating the global picture of how a collection of informal roles evolves across multiple timestamps is critical to understanding the overall patterns of role changes in the virtual community. For example, E3 commented that ``\textit{we can know which role occupies the majority of the game network over time}'' and ``\textit{how the minority gradually evolves into the majority,}'' said E4. Therefore, an intuitive visualization is demanded to achieve such analytical purposes.

\textbf{R5 Summarize interactions leading to informal role changes.} E3 -- E4 indicated that understanding the paths behind such informal role changes is critical for designing appropriate social mechanisms and guidelines. Given various interactions that can determine a player's socialization and engagement that result in a player's informal role, our approach should provide a clear role change trajectory and details of actions at both cluster and individual levels.

\section{Overview of RoleSeer}

\par Based on the requirements mentioned above, we propose a visual analytics system named \textit{RoleSeer} to support domain experts to better inspect informal roles and patterns behind their dynamics in the context of a given \textit{MMORPG}. \autoref{fig:pipeline} demonstrates the system architecture and pipeline. First, \textit{RoleSeer} processes in-game interaction logs from the \textit{MMORPG} and constructs undirected graphs to denote players' interactions and relationships from raw data. These undirected graphs are then fed into the back-end engine, which uses the proposed graph embedding model and clustering module to capture structural identity and infer informal roles (i.e., player cluster) (\textbf{R1}). To investigate and understand the potential informal roles, multi-level visualizations are integrated into \textit{RoleSeer} to fulfill the requirements from a coarse level to a fine-grained level. The front-end visualization consists of a \textit{Network Overview}, a \textit{Role Transition Overview}, a \textit{Role Transition Projection}, an \textit{Event Interaction View}, and an \textit{Individual Interaction View}. The network overview enables domain experts to look through all snapshots and their metric statistic information, and the experts can quickly locate the snapshot of interest (\textbf{R2}). The role transition overview focuses on the interconversion and evolution of informal roles. The experts can observe the characteristics of each informal role and the transitions between a variety of informal roles in a comparative manner. In a specific informal role, the experts can figure out which cluster the players within the cluster come and go (\textbf{R3, R4}). The role transition projection demonstrates the metric changes of the corresponding players that experience role transitions. Finally, the event interaction view and individual interaction view jointly provide evidence from the interaction event sequence during role transition. Thus, the experts could infer the potential reasons behind the role transition and obtain rich insights (\textbf{R5}).

\section{Back-end Engine}
\par The back-end engine first constructs a dynamic interaction network that models social interactions among players in the showcased \textit{MMORPG}. Then, it generates the network embedding representation that captures the structural identity throughout players' dynamic gaming process. The resulting network embeddings are reduced to a low-dimensional space for node clustering to facilitate the detection and exploration of the potential informal roles.

\subsection{Dynamic Social Network Construction}
\par The players' dynamic interactions throughout the gameplay are depicted by a dynamic social network. Specifically, we model the dynamic social network as a sequence of $T$ snapshots \{$G_1, G_2, ..., G_T$\}. $T$ denotes the number of time steps. Each snapshot is an undirected graph $G_t$ = $(V_t, E_t, t)$, where $V_t$ represents the set of the existing or active players at the $t^{th}$ snapshot. The edges $E_t$ denotes social relation strength measured by the latest intimacy value between the players at $t$. The intimacy value between two players is accumulated when an interaction between them happens. As suggested by the game team that the average gameplay duration for an in-depth player in this MMORPG is around $6$ hours, we set the size of the associated time window to $6$h for each snapshot. To reduce the computation time for constructing the network, we only reserve active players within each snapshot, i.e., if a player is temporarily offline within $6$h, he/she would be excluded in the network because the player does not contribute to the network evolution. When he/she returns to the game, his/her interactions with the others and the corresponding intimacy would be restored. 

\subsection{Proposed Method and Baseline Models}

\par To infer potential informal roles, we need a latent representation of each node (i.e., each player) to capture its structural identity in the dynamic evolving network. We introduce three representative dynamic network embedding methods: 1) Diachronic Node Embedding (DNE)~\cite{xu2018exploring}, 2) an alignment version based on \textit{struc2vec}~\cite{ribeiro2017struc2vec}, i.e., $struc2vec_{align}$, and 3) \textit{dynAERNN}~\cite{goyal2020dyngraph2vec}. \textit{DNE} learns node embedding by modeling neighboring local information in a dynamic graph. However, it is incapable of learning structural identity due to the nature of \textit{deepwalk} used in \textit{DNE}, i.e., the structurally similar nodes that locate far apart will not have equal representation according to \textit{deepwalk}~\cite{perozzi2014deepwalk}. To eliminate the limitation, we replace \textit{deepwalk} with \textit{struc2vec}, which encodes node similarity at different scales and thus can have a global perspective. The proposed method is named $struc2vec_{align}$. Similar to \textit{DNE}, we attach a \textit{TransE}~\cite{bordes2013translating} module to \textit{struc2vec}, which aims to align the \textit{struc2vec} embeddings between the continuous networks. On the other hand, the state-of-art \textit{dynAERNN} proposes an encoder-decoder architecture to predict missing links in a dynamic graph, and the last hidden layer outputs the learned embeddings. After the embeddings are generated, we use the dimensionality reduction technique \textit{t-SNE}  to project the node embeddings generated from the above three approaches into a $2$-dimensional space. Then, we cluster the projections using \textit{X-Means}~\cite{pelleg2000xmeans}, an optimized clustering algorithm that extends \textit{K-Means} in terms of calculation efficiency and efficient estimation of the number of clusters. The resulting clusters are then considered as informal role candidates. In the following subsection, we first design evaluation metrics to compare and evaluate the node embeddings generated from the above three approaches and then explore the identified informal roles.

\subsection{Evaluation Metrics}
\par The purpose of node embedding and subsequent clustering is to identify the potential informal roles a player may experience over time. Thus, an informal role cluster should capture diverse temporal information, maintain the structural identity within a cluster, and maximize the inter-cluster difference. In other words, our evaluation needs to examine which intra-cluster node embedding method can better maximize the diversity of temporal information and maintain the structural identity (i.e., minimize the structural difference). We hence adopt the following local and global metrics~\cite{li2018embeddingvis}, including local metrics like 1) \textit{Degree}; 2) \textit{Leverage Centrality} that depicts neighboring relationships, and global metrics 3) \textit{Page Rank}; 4) \textit{Closeness}; 5) \textit{Within Module Degree}, and 6) \textit{Betweenness}. In addition, to capture the diversity of temporal information, we use 7) \textit{Time Distribution} as a statistical metric to measure the intra-cluster temporal diversity. To calculate the intra- and inter-cluster differences in terms of the above metrics, we adopt an Inter-Cluster Metric as 

\begin{equation}
    \begin{aligned}
        \label{equ:inter_cluster}
        \frac{1}{\|S\|}\sum_{\substack{(p,q)\in S,\\ p \neq q}} KL\_Divergence(hist(Met_i(p)), hist(Met_i(q)))
    \end{aligned}
\end{equation} to compute the mean of KL Divergence, where $S$ is the set of the clusters, $hist(\cdot)$ denotes the histogram of the normalized $i^{th}$ metric. $Met_i(\cdot)$ outputs a vector containing the $i^{th}$ metric of all nodes. We use an \textit{Intra-Cluster Metric} as 
\begin{equation}
    \label{equ:intra_cluster}
    \frac{1}{\|S\|} \sum_{p\in S} var(Met_i(p))
\end{equation} to measure the mean of variances of metrics within one cluster, where $var(x)$ represents the variance of $x$.

\begin{table*}[h]
\caption{Inter-cluster metrics for different models (the larger, the better)}
  \vspace{-3mm}
\label{tab:inter}
\centering \small
\begin{tabular}{c|c|c|c|c|c|c}
\hline \hline
Model     & Degree      & PageRank    & Betweenness & Leverge Centrality & Within Module Degree & Closeness \\ \hline \hline
$DNE$       & 0.821670504          & 0.411103418          & 0.140934534          & 0.16379114           & \textbf{0.780515731} & 1.369339797          \\
$dynAERNN$ & 0.796215423          & 0.398164498          & 0.459215495          & 0.377308901          & 0.32386315           & 0.712463516          \\
$struc2vec_{align}$ & \textbf{3.366409078} & \textbf{1.298493177} & \textbf{0.940996241} & \textbf{1.314054328} & 0.586676757          & \textbf{1.681372761} \\ \hline
\end{tabular}
\end{table*}

\begin{table*}[h]
\caption{Intra-cluster metrics for different models (the smaller, the better)}
  \vspace{-3mm}
\label{tab:intra}
\centering \small
\begin{tabular}{c|c|c|c|c|c|c}
\hline \hline
Model     & Degree      & PageRank    & Betweenness & Leverge Centrality & Within Module Degree & Closenes \\ \hline \hline
$DNE$       & \textbf{0.004741896} & 0.001819743          & \textbf{0.00066172} & 0.026901872          & \textbf{0.004532516} & \textbf{0.012460878} \\
$dynAERNN$ & 0.00917859           & \textbf{0.001512473} & 0.000947724         & 0.022042176          & 0.008951476          & 0.01330977           \\
$struc2vec_{align}$  & 0.007649969          & 0.002744551          & 0.002403099         & \textbf{0.021968981} & 0.010843563          & 0.012954739          \\ \hline
\end{tabular}
\end{table*}

\subsection{Result Analysis}
\par
\begin{figure}[h]
 \centering 
 \includegraphics[width=0.48\textwidth]{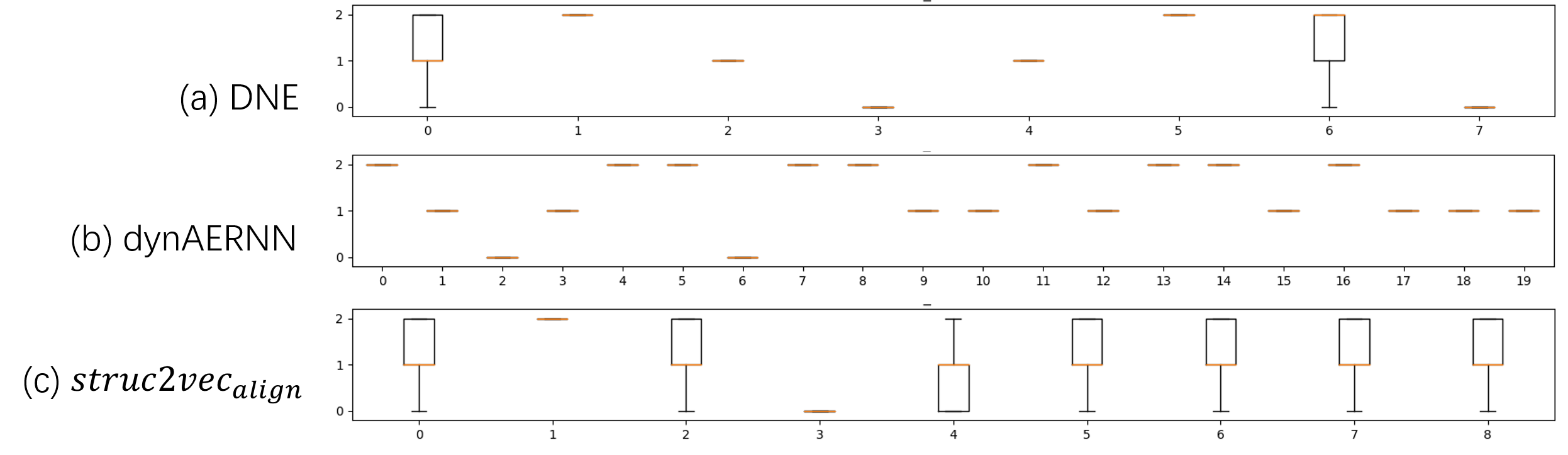}
  \vspace{-3mm}
 \caption{Time distribution over clusters, where the orange line denotes the median. The time distribution within each cluster in \textit{dynAERNN} is very concentrated, implying that \textit{dynAERNN} does not maintain the diversity of temporal information well. On the contrary, the distribution in the other two methods, are more diverse within one cluster, particularly, $struc2vec_{align}$ achieves the best differentiation in terms of time distribution.}
 \label{fig:box-plot}
   \vspace{-3mm}
\end{figure}

 It is infeasible to evaluate the entire set of players due to the high computation complexity. Therefore, we thereby randomly sample a set of players within $3$-time snapshots (i.e., $18$ hours in total). We then unify all the generated network embeddings from the three methods as $128$-dimensional vectors to calculate the abovementioned metrics. \autoref{fig:box-plot} shows the fine-grained box-plot statistics of time distribution from the three methods. The time distribution within each cluster is very concentrated, i.e., only comprising players from one time snapshot, implying that \textit{dynAERNN} does not maintain the diversity of temporal information well. On the contrary, the distribution in the other two methods, i.e., \textit{DNE} and $struc2vec_{align}$, are more diverse within one cluster, particularly the result in \autoref{fig:box-plot} (c) achieves the best differentiation in terms of time distribution. To compare the metric performance more quantitatively, we demonstrate the inter-cluster and intra-cluster metric results in \autoref{tab:inter} and \autoref{tab:intra}, respectively. It is observed that most inter-cluster metrics of $struc2vec_{align}$ are higher than those in the other two methods, especially the two local metrics, i.e., \textit{Degree} ($3.366409078$) and \textit{Leverage Centrality} ($1.314054328$). In the intra-cluster metrics, the difference among the three models is relatively trivial compared to the gaps in \autoref{tab:inter}. Although \textit{DNE} wins in most metrics, $struc2vec_{align}$ still shows a comparable performance. Furthermore, the performances in both the inter- and intra-cluster metrics are similar to the conclusion attained by \textit{EmbeddingVis}~\cite{li2018embeddingvis}, which confirms that \textit{struc2vec}-based embeddings preserve \textit{Degree} better than the \textit{deepwalk}-based methods. To sum up, the experiments demonstrate that our proposed $struc2vec_{align}$ can well maintain structural identity and capture diverse temporal information over the dynamic network. Therefore, we use the network embeddings generated by $struc2vec_{align}$ to identify the potential informal roles.

\section{Front-end Visualization}
\par
\begin{figure*}[h]
 \centering 
 \includegraphics[width=\textwidth]{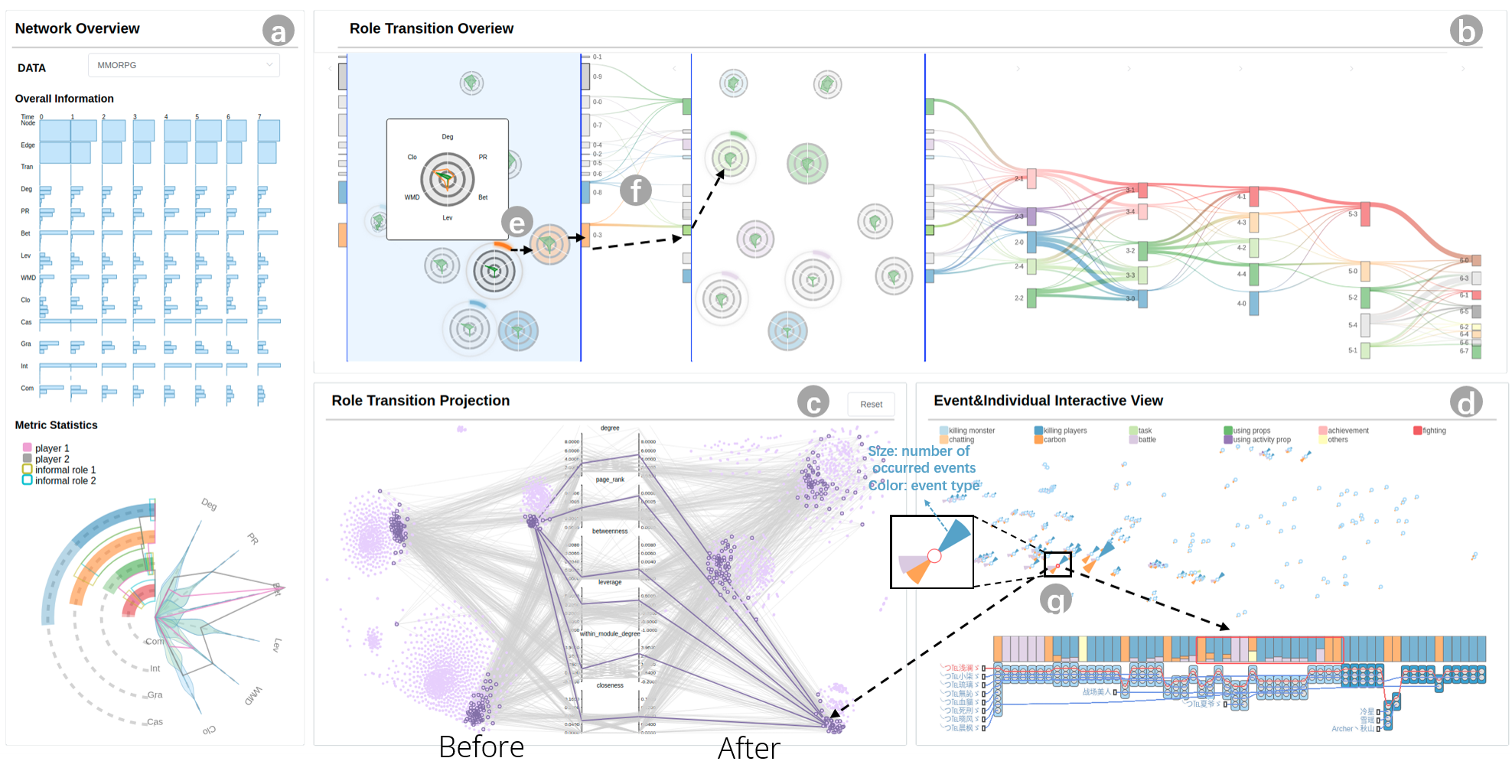}
  \vspace{-6mm}
 \caption{System interface of \textit{RoleSeer}. (a) The network overview supports data selection and summarizes the metric distribution across multiple snapshots. The metric statistics comprise four in-game metrics (left) and six graph metrics (right). (b) The role transition overview (i.e., unexpanded snapshot) shows the overall interconversion and evolution across snapshots (f), and the expanded snapshot presents the role transition within a snapshot (e). (c) The role transition projection projects the graph embedding results of players within selected informal role clusters. The players with transition are highlighted by grey lines and linked across two informal roles. Metric changes are also visualized in the middle axis between the two projections. The selected player's metric changes are highlighted (g). (d) The event and individual interaction view shows the projection results of players' event patterns and narrates the selected ego's interaction with other alters, as well as the percentage of each event.}
 \label{fig:role_transition_overview}
   \vspace{-3mm}
\end{figure*}

 \textit{RoleSeer} augments familiar visual metaphors to enable domain experts to focus on analysis. We strictly follow the mantra ``overview first, zoom and filter, then details-on-demand''~\cite{shneiderman2003eyes} to guide users in exploring the formation, interconversion, and transition of informal roles in the context of an MMORPG. Based on these principles and the preceding requirements, we develop four visualizations that allow the output of the back-end model to be quickly inspected at the overview, role cluster, and individual levels. Specifically, we design a \textit{network overview} (\autoref{fig:role_transition_overview}(a)) to locate the snapshot of interest and observe its characteristics in terms of various graph metrics; a \textit{role transition overview} (\autoref{fig:role_transition_overview}(b)) and a \textit{projection view} (\autoref{fig:role_transition_overview}(c)) to assist domain experts in obtaining an overview of the characteristics of the potential informal roles and the transitions among them; an \textit{event interaction view} to explore the reasons behind role changes, and an \textit{individual interaction view} to visualize individual's interactions with other counterparts (\autoref{fig:role_transition_overview}(d)).

\subsection{Role Transition Overview}
\par The role transition overview is designed for three purposes. First, appropriate metrics should be selected to represent each detected informal role cluster (P1). The clusters are generated by the \textit{X-Means} algorithm that accepts the output of the embedding vectors generated by the $struc2vec_{align}$ mentioned above. However, interpreting these ``clusters'' from the perspective of the embedding vector is rather abstract because the dimensions in a typical embedding do not have a particular meaning~\cite{smilkov2016embedding}. Instead, the experts are more familiar with specific metrics such as \textit{PageRank} and \textit{betweenness}. Therefore, showing appropriate metrics representing each identified informal role cluster can help domain experts characterize and understand the role clusters. Second, the difference between the role clusters should be presented to facilitate the exploration and understanding of the cluster. Visualizing the characteristics of each role cluster is necessary for evaluating that the entities in the cluster are collectively gathered in a meaningful manner. Precisely, metric distributions of and differences between multiple role clusters should be placed in the same context, allowing domain experts to compare multiple role clusters simultaneously (P2). Third, the interconversion and transition among the role clusters should be intuitively visualized to enable domain experts to understand the dynamic changes of informal roles. Particularly, tracking the evolution of informal roles and how players shift their positions is necessary for understanding the evolving gaming experiences of different players. Thus, an intuitive way to visualize the interconversion and transition among the roles is desired (P3).

\begin{figure}[h]
 \centering 
 \includegraphics[width=0.48\textwidth]{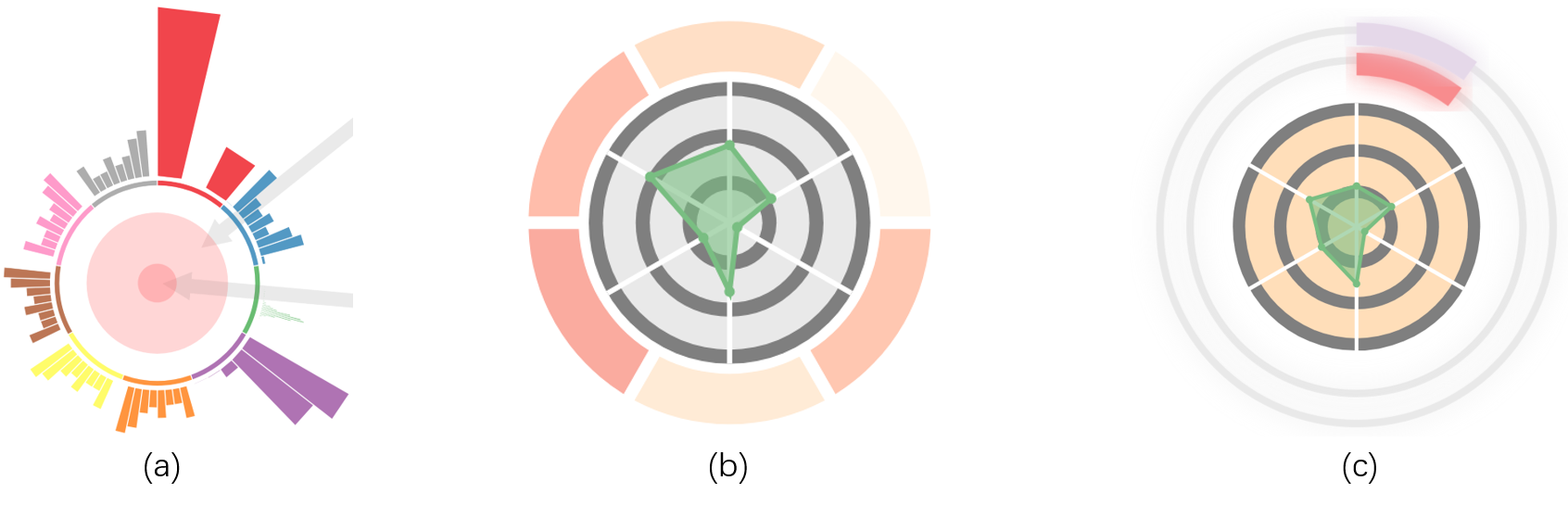}
  \vspace{-6mm}
 \caption{Design alternatives of informal role clusters. (a) uses bar height to encode metric distribution and arrows to denote an interconversion between two roles; (b) uses colors of the outer circle to encode metrics and inner radar chart to indicate interconversion from other informal roles; (c) takes radar chart to represent the metrics and "planet rings" to tell the interconversion to different informal roles.}
 \label{fig:alternatives}
   \vspace{-3mm}
 \end{figure}

\textbf{Glyph Design.} To meet the first and the third purpose (P1, P3), we design a novel glyph to represent the informal role cluster, consisting of an inner circle and an outer circle. Inspired by~\cite{Tsang20TradAO}, the inner circle is a Polar coordinate where each white axis represents one metric, and the color of the inner circle corresponds to a unique informal role. The glyph size is determined by the player amount holding the informal role. We obtain its average value for each metric and connect each metric to form a radar chart that characterizes the informal role. The outer circle, i.e., a ``planet ring,'' depicts the interconversion between informal roles within a snapshot. We define that in snapshot $T$ (containing timestamps from $t_i$ to $t_{i+N}$ and $N$ is the period of the snapshot), and interconversion between informal roles occurs when a player switches its position from informal role $A$ at timestamp $t_i$ ($i<N$) to another informal role $B$ at timestamp ($t_{i+1}$). For example, in \autoref{fig:alternatives}(c), each colored outer circle around the inner circle indicates a transition, of which the color denotes the target role cluster of the transition. The transition ratio (i.e., the amount of interconversion players divided by the amount of all players in the informal role) corresponds to the clockwise arc angle of the ring. To meet the second purpose (P2), we follow a classic comparative visualization design, namely, the superposition proposed by Gleicher et al.~\cite{gleicher2011visual} to visually compare two informal roles. Notably, we display an enlarged glyph without the outer circle when users hover on a particular glyph, as shown in ~\autoref{fig:alternatives}(c). A legend of metrics is labeled around the radar chart. Two radar charts, i.e., the source and the destination informal role, are stacked on the inner circle to enable a straightforward comparison.

\textbf{Design Alternative of the Glyph.} The final glyph undergoes several user-centric design iterations. As shown in \autoref{fig:alternatives}, we initially designed a circular grouped bar glyph to represent the role cluster. Each segment represents one metric, and the bar charts on the top of each segment indicate the value distribution of that metric. The size of the glyph represents the number of players holding the role. The size of the red circle on the glyph center indicates the number of players shifting their positions. Our experts commented that this design did not use the inner space of the circle efficiently, and the metric distribution is hard to compare with each other.

\par \autoref{fig:alternatives}(b) borrows the design in~\cite{Tsang20TradAO}, where the color shade of the outer ring denotes metric values, and the inner circle represents the number of players transferring to this role cluster. This design is more spatially efficient but suffers from an ineffective comparison of metrics due to human's insensitivity to colors. Our experts finally chose the last version that appends interconversion to each informal role, avoiding drawing any connection among role clusters. Furthermore, the radar chart is much more distinguishable.

\begin{figure*}[h]
 \centering 
 \includegraphics[width=\textwidth]{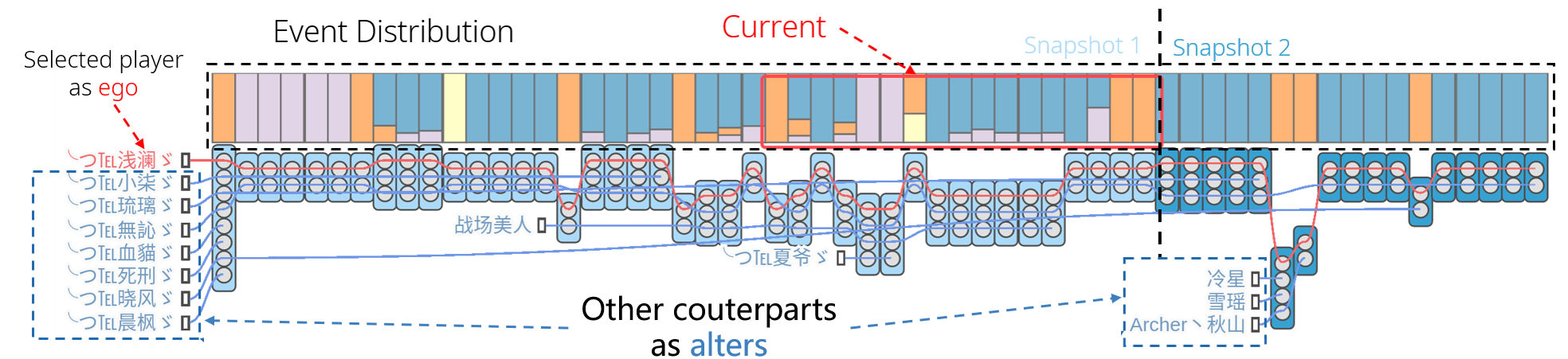}
  \vspace{-6mm}
 \caption{The individual interaction view visualizes how an ego interacts with ``1-degree'' alters over the course of his/her gaming process. The event distribution is shown by stacked bars and the interaction among player is supported by a storyline-based metaphor. The red rectangle indicates the current events corresponding to the selected player.}
 \label{fig:individual}
   \vspace{-3mm}
\end{figure*}

\par \textbf{Transition Flow across Snapshots.} While the role transition glyph conveys the potential role shifting within a shorter period (i.e., $18$h), tracking the long-term interconversion and evolution is also important. To this end, we embed a transition flow in the role transition overview. First, we find out the identity of each informal role and its transition across snapshots. Specifically, we use the Jensen-Shannon Divergence (JS)~\cite{manning1999foundations} to measure metric distribution similarity between two informal roles as follows:
\begin{equation}
JS(p||q) = \frac{1}{2}*KL(p||\frac{(p+q)}{2}) + \frac{1}{2}*KL(q||\frac{(p+q)}{2}),
\end{equation}
where $KL$ is the KL Divergence and $p,q$ are the distributions to be measured (same meaning as in \autoref{equ:inter_cluster}). It provides a symmetric measurement for players' metrics. For each informal role $i$, we consider the role $j$ with the least JS value as $i$'s most similar informal role and depict the two informal roles ($i$ and $j$) using the same color. For the case where the least JS value between $i$ and $j$ is significantly large (e.g., remarkably larger than a threshold), we consider that no existing informal role $j$ is similar to the current role $i$. Thus, we just assign a new color to the informal role $i$. We align players’ IDs to track their transition from an earlier informal role to a latter one across two subsequent snapshots. As shown in \autoref{fig:role_transition_overview}, we use a Sankey diagram to present identity transition across multiple snapshots. Each column stands for a snapshot. Each node denotes an informal role within the snapshot, and each flow represents a transition between informal roles across subsequent snapshots. By integrating the Sankey diagram with the above glyphs, the subtle role change occurring within a snapshot and a significant role change happening across multiple snapshots can be simultaneously witnessed, allowing domain experts to interactively identify the role change within one snapshot and track consistent informal roles (i.e., rectangles with the same color) across continuous snapshots.

\subsection{Role Transition Projection}
\par Role transition projection (\autoref{fig:role_transition_overview}(c)) looks into the players' position within a specific informal role in detail to analyze interconversion and evolution at the informal role level. It supports the comparison between informal roles in a fine-grained manner. We project the embedding vector of each player onto a 2D space using \textit{t-SNE} projection. We choose \textit{t-SNE} as the dimensionality reduction technique since it demonstrates superiority in producing a 2D projection that reveals meaningful insights about data such as clusters and outliers~\cite{kim2016pixelsne}. Also, it is more visually interpretable than that eigenanalysis and more intuitive than multidimensional scaling (MDS) depending on the distribution~\cite{li2018embeddingvis}. Between the two spaces of the informal roles, we connect all nodes to their metrics in the central axes, similar to the design of the parallel coordinate plot (PCP). Purple circles would highlight the players occurring role changes, and their metric changes could be easily witnessed through the axis values. Users can lasso on the role transition projection view to select a specific subset of players, and the corresponding content in the event interaction view will be updated.

\subsection{Event Interaction View}
\par The event interaction view facilitates exploring players' event sequences that lead to specific role changes. Specifically, this view helps identify the particular event(s) that explain the interconversion and transition among the roles. In reality, the length of players' event sequences within a period may vary significantly. We compress those consecutive events into one event to decrease the sequence length. Then, the compressed event sequences are fed into the \textit{Doc2Vec} model~\cite{le2014distributed} to generate the latent representations. We project the latent representations of all event sequences corresponding to the selected role change by \textit{t-SNE} to visualize the fine-grained patterns that drive the role change intuitively. In \autoref{fig:role_transition_overview}(g), each dot represents one event sequence, and their placements reflect their relative similarities to help domain experts discover patterns. We further enhance the event interaction view with a coxcomb glyph due to the effectiveness of glyphs in facilitating visual comparison and pattern recognition to better identify the differences between clusters and find representative event sequence dots. Particularly, we use event distribution to differentiate and characterize the dots. Accordingly, the outer sectors in our coxcomb glyph design represent the number of each event. However, the glyph design may bring about a visual clutter when many dots exist. Similar to the strategy in~\cite{zhao2017skylens}, to alleviate this issue, we decrease the glyph opacity in default to ensure that individual glyphs can be quickly inspected. When domain experts hover over one glyph, it would be brought to the foreground. We also provide interactions like panning and zooming for focusing on a particular region. Regarding design alternatives, we did not select classic star/radar glyphs to encode the event number distribution because the number of events is not always the same for each event sequence pattern. Meanwhile, the line in the star/radar glyphs is challenging to perceive when the color saturation is low and the glyphs are small.

\subsection{Individual Interaction View}
\par To provide a detailed inspection and explanation of event patterns, the individual interaction view summarizes and visualizes individual players' interaction gaming process with other counterparts (\textbf{R5}). The input of the individual interaction view includes the selected player and his/her friends in all interaction rounds.

\par \textbf{Visual Encoding for Interactive Events.} We harness a stacked bar chart and arrange the stacked bars over time to visualize the percentage of each event in a particular period, i.e., $10$ minutes, demonstrating how an individual player interacts with other counterparts. Specifically, the selected player is considered ``ego'', and other counterparts are ``alters'', forming an ``ego-alters'' social network. Initially, we compared the stacked bar design with the pie charts. However, due to the dense space between two subsequent periods and the fact that there are no considerable differences in the ratio of different events, it is hard to obtain a precise distribution on a small pie. Therefore, we finally chose the stacked bar design.

\par \textbf{Visual Encoding for Individual Interaction.} In the individual interaction view, we align each interaction process (i.e., $10$-minute time period) along the $x$-axis and show all interactions within two-time snapshots (i.e., $2 \times 18$-hour time period). However, determining the coordinates along the x-axis of each interaction round is easy, while placing the $y$-coordinates of each interaction round to minimize the number of potential link crossing is an NP-hard problem and is thus difficult to obtain an exact solution. We, therefore, use a heuristics approach following the idea of narrative visualization~\cite{li2017visual,munroe2009movie}. Suppose we have a series of interaction rounds representing them as rounded rectangles (\autoref{fig:individual}). In each rectangle, circles indicate the players that participate in this round, and the background color refers to the snapshot when events happen. Precisely, we first align the interaction rounds in the $x$-direction according to the round id (i.e., the order of appearance). The $y$-coordinate of each interaction round is adjusted to minimize the overall curvatures and intersections when linking identical players across consecutive interaction rounds. Specifically, the $y$-coordinate in one interaction round lies within the $y$-range of the median of the players participating in this round. We sort the interaction rounds in the $y$-direction to ensure that the densest ones (the ones containing the most players) are as far away from each other as possible to avoid possible visual clutter and maximize the utilization of the available space. We denote all the players that participate in the interaction process as $P = p_0, ..., p_k$ with the index of $k$ defining the order of their appearance. We then calculate a temp-$y$-coordinate for each client as $temp_y(p_k) = Index(p_k) * space$, in which space is a constant value indicating a horizontal gap between two adjacent players. After determining all the positions of interaction rounds and each player in every interaction round, we visualize the players' participation as shown in \autoref{fig:individual}. We link identical players across subsequent interaction rounds by curves.

\subsection{Metric Statistics}
\par In the role transition view and role transition projection, we use abstract graph metrics to describe the characteristics of the detected informal roles. Although these metrics facilitate our exploration of the informal roles, they are hard to interpret from the perspective of game designs due to the lack of semantics. In other words, the experts cannot go further to find the shared game characteristics of informal roles (\textbf{R5}). To address this concern, we design a novel metric statistics glyph to include in-game statistics in the same context. Specifically, the in-game metrics include \textit{cash (Cas)}, \textit{grade (Gra)}, \textit{intimacy (Int)}, and \textit{combat score (Com)}, and the glyph summarizes both the informal role's information and a specific player's information as follows.

\begin{figure}[h]
\vspace{-3mm}
 \centering 
 \includegraphics[width=0.5\textwidth]{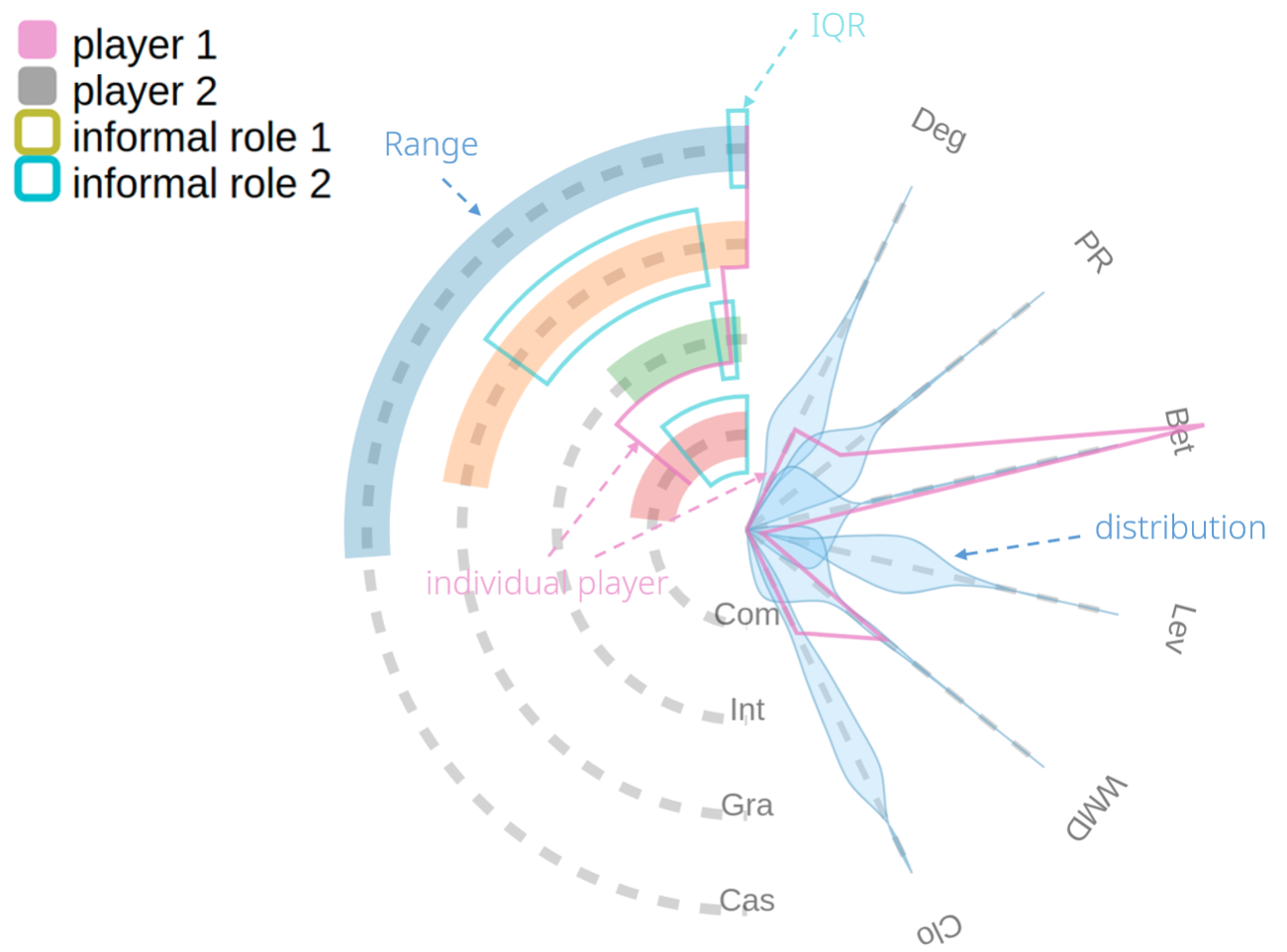}
  \vspace{-6mm}
 \caption{Metric statistics. The left side draws in-game metrics by a ``box plot'' along the circles, where the filled ring denotes the range and the hollow arc represents the interquantile range (IQR). The right side shows the distribution of abstract graph metrics on axes. The information of an individual player is also drawn by a link (left) and a polygon (right). This view supports a comparison between two selected informal roles and individual players.}
 \label{fig:metric_statistic}
   \vspace{-3mm}
\end{figure}

\begin{figure*}[h]
 \centering 
 \includegraphics[width=\textwidth]{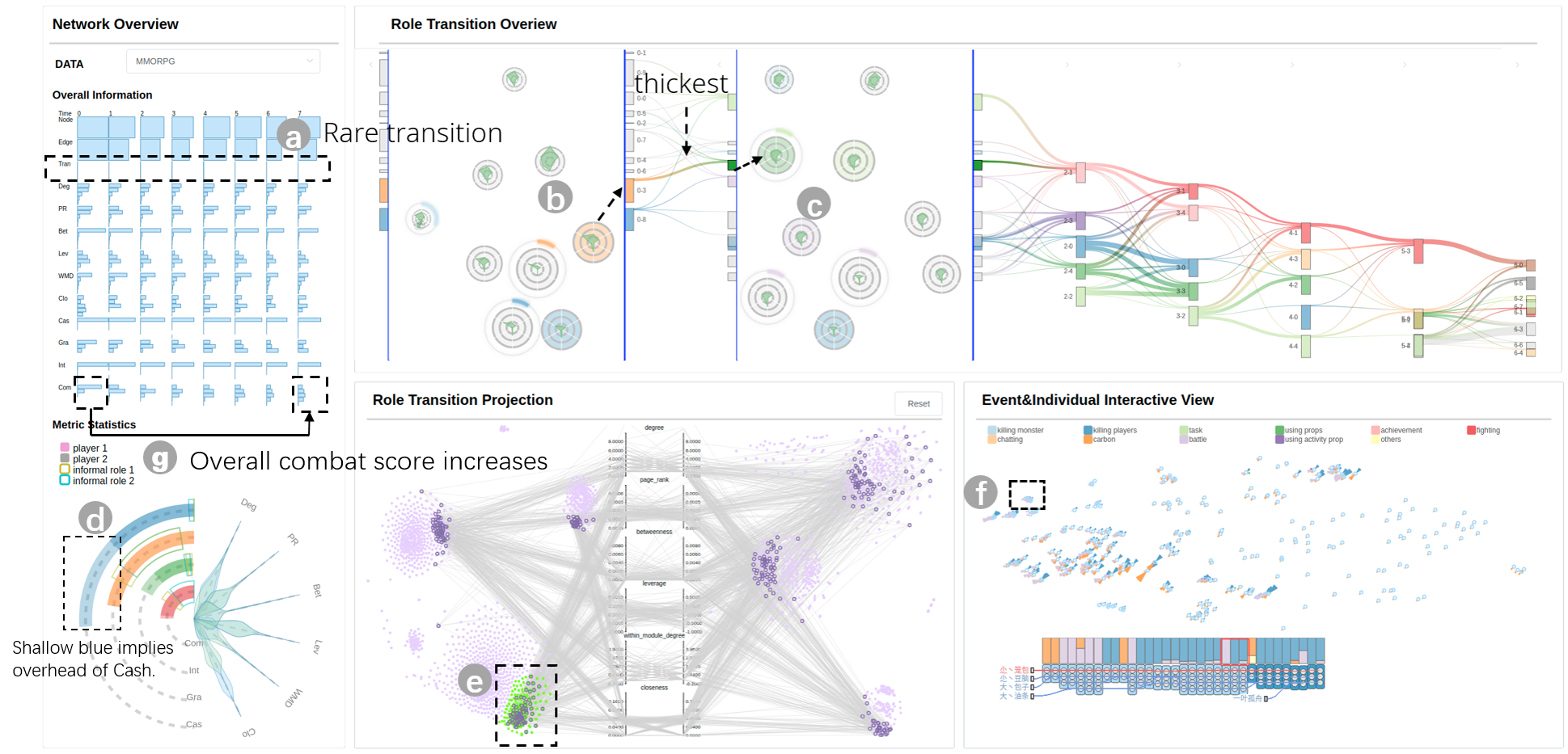}
  \vspace{-6mm}
 \caption{(a) shows the low transition percentage and (g) shows that players' overall \textit{combat score} increases over time. (b) and (c) are the beginning and ending of a transition. Their metrics are shown by a radar chart. In (d), in-game metrics and abstract graph metrics of two informal roles are visualized and compared. (e) is the lassoed dots and the corresponding players are shown in the event interaction view. (f) are aggregated dots where players show similar event patterns.}
 \label{fig:case10}
   \vspace{-3mm}
\end{figure*}

\par \textbf{Informal Role's Information.} Each dashed ring denotes a metric, where a box plot presents the distribution of players' information on that metric. The color ring stands for the data range, and the hollow arc represents the box plot's interquartile range (IQR). In this way, the region of the stacked color ring has a deeper color, and the difference between two IQRs can be observed, which is convenient to capture the similarity between the ranges and the difference between the two data distributions. At the right side of \autoref{fig:metric_statistic}, a radar chart shows the distribution of an informal role's metrics. The area along the axis shows the metric distribution.

\par \textbf{Individual's Information.} We propose different designs on the left and right sides for an individual player. In the left circles, a line in the rings indicates its value. We connect them by arcs. In the right chart, a polygon is constructed by connecting metric values on all axis and the center of the circle. Therefore, a significant visual difference can reflect data differences in one dimension.

\subsection{Interaction Among the Views}
\par Apart from the most defining capabilities of \textit{RoleSeer}, rich interactions are integrated to catalyze an efficient in-depth analysis, following the design mantra ``overview first, zoom and filter, then details-on-demand''. First, the general information in the \textit{network overview} helps users identify time segments of interest. Otherwise, they may get lost by the unexpanded columns in the \textit{role transition overview}. Then users can move to the specific timestamp, and the \textit{role transition overview} allows to expand multiple columns: 1) within a timestamp, users can \textbf{hover} a glyph to observe a cluster-level metric statistics, or \textbf{click} a glyph to move to the individual-level \textit{event \& individual view}; 2) across timestamps, users can \textbf{click} a link between columns to move to the individual-level \textit{role transition projection}, in which lassoing nodes shows the graph metric transformation in a group level. Furthermore, the \textit{metric statistics} show the game metric statistics at a group level. In the \textit{event \& individual interactive view}, \textbf{clicking} a node shows the corresponding player's information from three perspectives in a \textbf{link + view} manner: 1) the \textit{event \& individual interactive view} shows the event narrative; 2) the \textit{role transition projection} shows the metric transformation; 3) the \textit{metric statistics} shows the information of game characteristics.

\begin{figure*}[h]
 \centering 
 \includegraphics[width=\textwidth]{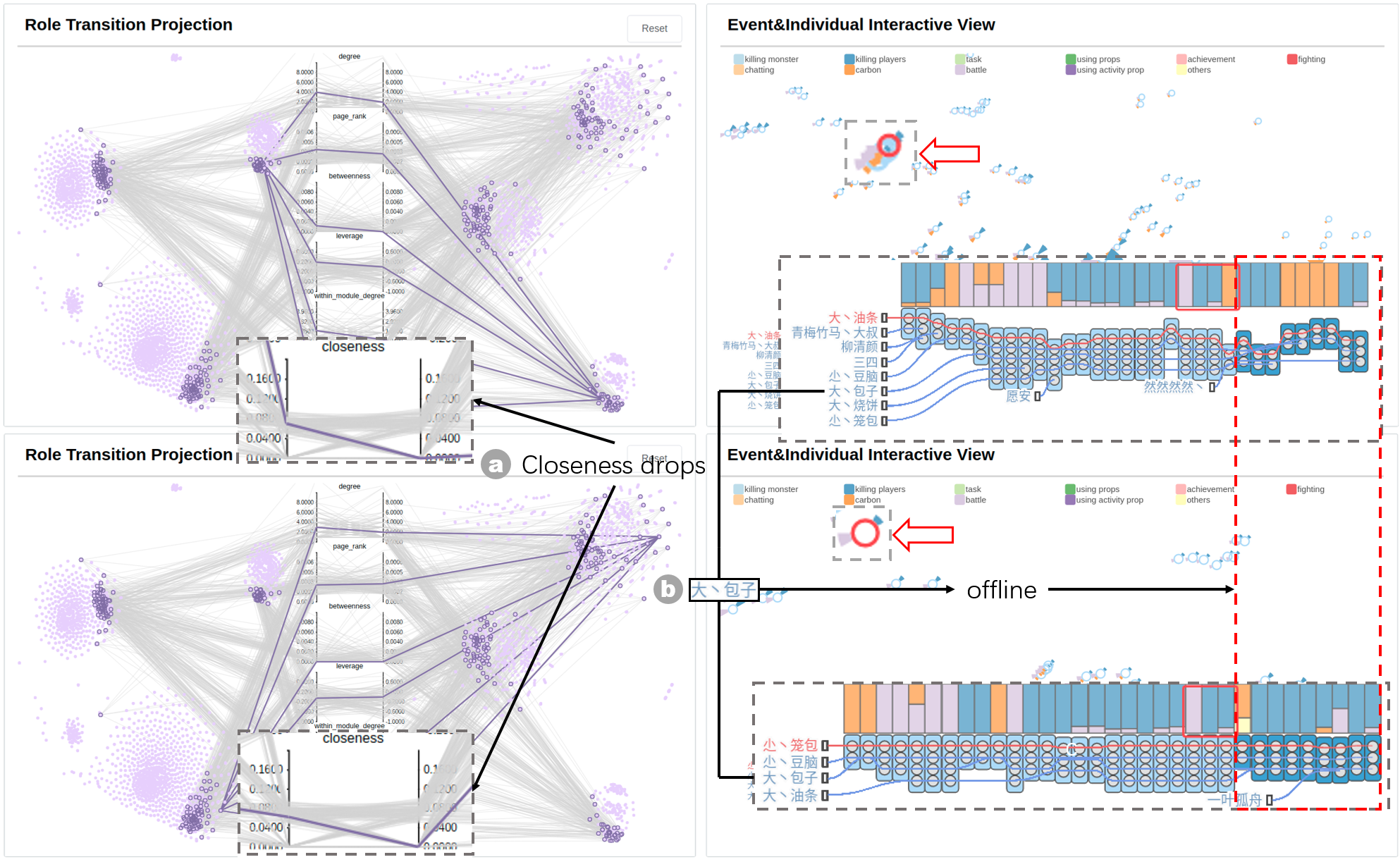}
  \vspace{-6mm}
 \caption{From the perspectives of different egos (red arrows) in one community, we observe a ``closeness'' drop (a) when the specific player is offline (b).}
 \label{fig:case1}
   \vspace{-3mm}
\end{figure*}

\section{Evaluation}
\par We evaluated the effectiveness of \textit{RoleSeer} in multiple ways from the visualization community~\cite{6634108,16041}. First, we described two case studies with our domain experts (E1 -- E4) who had participated in our user-centric design process. Second, we invited $12$ participants who had no exposure to our system and conducted a user study to further assess the potency of \textit{RoleSeer}. Finally, we conduct expert interviews about their user experience and feedback to \textit{RoleSeer}.

\begin{figure*}[h]
 \centering 
 \includegraphics[width=\textwidth]{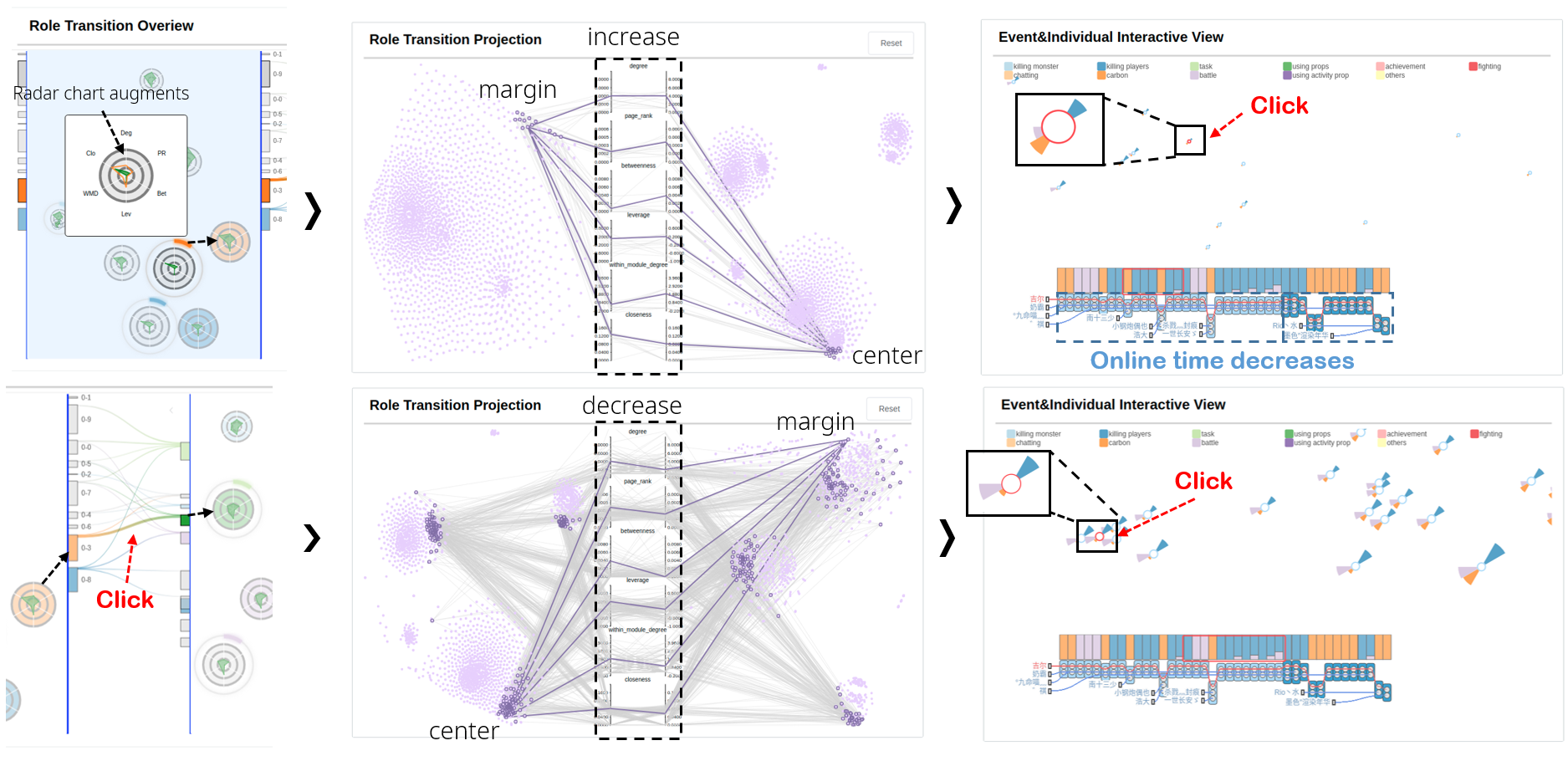}
  \vspace{-6mm}
 \caption{Within a continuous transition, one player at different snapshots is found. In the role transition projection, there is a positive correlation between transition direction and metric shift. In the event interactive view, the event distribution indicates that \textit{carbon} and \textit{battle} leads to the different transitions. In the individual interactive view, a shorter online time in the next time snapshot is observed.}
 \label{fig:case21}
   \vspace{-3mm}
\end{figure*}

\subsection{Case \uppercase\expandafter{\romannumeral1}: Identifying Players with Key Informal Roles}
\par The following activities occurred when our collaborating experts analyzed the players' social behaviors to uncover an interesting social role a particular player holds in the specific MMORPG.

\par \textbf{Observing the overall stability of the entire social network.} Our data analyst E2 first moved to the network overview to get a general idea of the whole of social network evolution status (\autoref{fig:case10}). He inspected the statistics information in terms of all transition (\textit{Tran}) percentages, the number of nodes and edges to identify the snapshot that significant drops and increases occur, as well as the distributions of graph metrics across all the snapshots, including \textit{degree (Deg)}, \textit{page rank (PR)}, \textit{betweenness (Bet)}, \textit{leverage centrality (Lev)}, \textit{within module degree (WMD)}, and \textit{closeness (Clo)}, adopted from the literature~\cite{li2018embeddingvis}. He witnessed that the transitions rarely occurred at each snapshot. Except for the snapshot ``0 -- 1'', he could not find a significant visual difference in terms of the online players (i.e., nodes) and their interactions (i.e., edges). He explained that at the initial release of the game, some players were just attracted by the novelty but may be away from the keyboard after a shallow gameplay taste. For the remaining metrics, he found they changed as expected. For example, in the beginning, the distribution of \textit{combat score} (Com) was mainly concentrated in smaller values. Later, the proportion of higher \textit{combat score} became higher (\autoref{fig:case10} (g)). He concluded that the game runs stably as expected.

\par \textbf{First impression on the identified informal roles.} E2 was then curious about our approach's identified informal roles, so he moved to the role transition overview. After a general observation of the view, he was pretty interested in the most significant nodes and edges in the first snapshot, so he compared the transition flows between the first and the second columns in the role transition overview. The top flow has the most instances that experience the transition, so he clicked on and expanded the space between the two columns next to the transition. Then, he located the two informal roles involving this transition and inspected them in detail regarding their metric distributions (\autoref{fig:case10}(b)(c)). By comparing the radar charts within two glyphs, he implies that (b) plays a more critical role than (c). Moreover, he clicks on (b) and (c) in order and the metrics statistic updates. In \autoref{fig:case10}(d), the advantage of (b) in \textit{grade} and \textit{combat scores} illustrates that players have more combat experiences and capabilities. In the blue arc of cash in \autoref{fig:case10}(d), he could see a shallow region, implying the intensive economic activities of the corresponding players.

\par \textbf{Exploring and understanding a transition.} Inspired by the glyph that presents an obvious transition pattern as shown in ~\autoref{fig:case10}, E2 clicked on the two corresponding informal roles of interest, the transition link between them, and the system updated the role transition projection and event interaction view. The corresponding players and their interaction events were displayed. Notably, E2 observed three sub-clusters within the informal role (b), and the deep purple dots highlighted correspond to the players who experienced the role transition. E2 found that most players moved from the cluster periphery in the informal role (b) to more centric positions in the informal role (c). He further lassoed one part of the dots (\autoref{fig:case10}(e)), highlighting the corresponding event interaction glyphs. From the event interaction glyphs, he identifies a collection of dots in which nodes within have a similar visual appearance, and he just clicked on it one by one (\autoref{fig:case10}(f)). The corresponding dot was also highlighted in the role transition projection view, as shown by dots connected by the deep purple links, facilitating tracking the player from its position within the informal role (b) to the metric value on the left axis to another value on the corresponding right metric axis, and finally to the position within the informal role (c). Furthermore, in the individual interaction view, the detailed interactions of the player with its egocentric community were shown so that E2 could observe how the player interacted with his/her alters.

\par \textbf{Identifying players holding key informal roles.} In the individual interaction view, E2 guessed that the ego and some alters were offline friends since they shared a similar formation of their game nicknames (\autoref{fig:case10}(f)). From the metric distribution of the ego (upper part in \autoref{fig:case1}(a)), E2 found that its degree decreases a bit. Still, the closeness drops significantly, which was caused by the offline event of one player (P) shown by the interaction event sequence in the individual interaction view (upper part in \autoref{fig:case1}(b)). Therefore, E2 proposed a hypothesis that the absence of P leads to the drop of the ego's closeness, and P held the informal role that maintains the closeness of the community. To further validate this hypothesis, E2 explored another sub-cluster in the role transition projection and identified another alter of the community (red arrow in \autoref{fig:case1}(b)). The metric change of this alter was similar to the previous ego (lower part in \autoref{fig:case1}(a)), and in its individual interaction view, P was offline when the shift in closeness metric occurred (lower part in \autoref{fig:case1}(b)). Thus, E2 concluded that P played the informal role that preserves community closeness. ``\textit{I am quite inspired by this finding, as I only need to track this player who can maintain the stability of the community.}'' Following the same procedure, E2 only needed to identify and study the players holding a similar informal role with P to ensure the stability of a particular game community.

\subsection{Case \uppercase\expandafter{\romannumeral2}: Revealing Reasons Behind Role Transitions}
This case demonstrates the efficiency of \textit{RoleSeer} for tracking informal role transitions and understanding the underlying reasons behind such shifts.

\par \textbf{Identifying players experiencing transitions from community periphery to center.} E1 first took a look at the role transition overview and expanded the first columns, i.e., the first snapshot. He compared two informal roles and found a significant difference regarding its metric distribution, i.e., the metrics after role transitions have an overall augment in all dimensions. For further inspecting the details, E1 clicked on the two glyphs and the transition ring to visualize more information in the role transition projection and event interaction views. Particularly, in the role transition projection view, he discovered that all players with transitions went from the margin areas to the more centric regions. Correspondingly, all metric distributions except for closeness rise to a higher level or stay unchanged. Therefore, E1 proposed a hypothesis that when a player's network metrics drop or increase after a role transition, he/she shall shift the role to margin or center areas.

\begin{figure*}[h]
 \centering 
 \includegraphics[width=\textwidth]{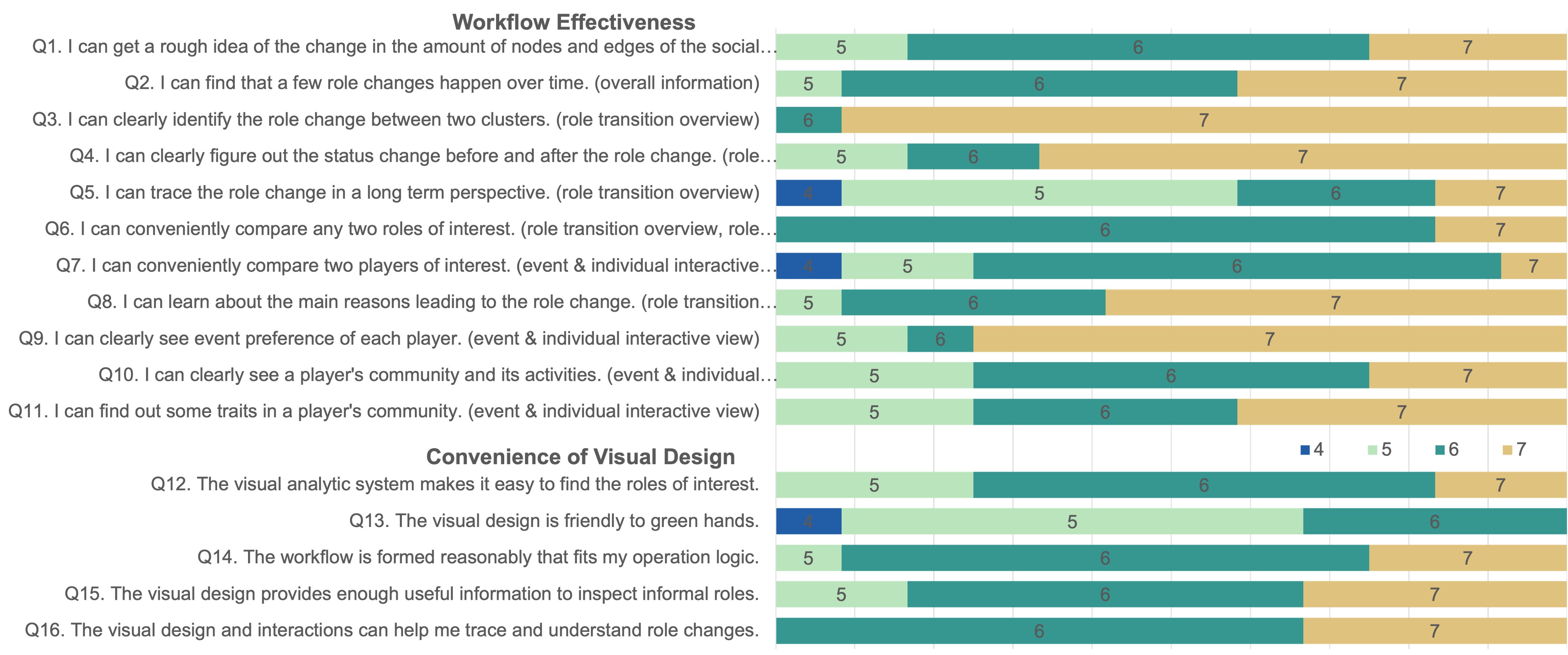}
  \vspace{-6mm}
 \caption{The results about users' perceptions on \textit{RoleSeer} regarding its workflow effectiveness and the convenience of visual design. $1 - 7$ represents ``strongly disagree'' to ``strongly agree'' on each statement (Q1 -- Q16).}
 \label{fig:user_study}
   \vspace{-3mm}
\end{figure*}

\par Following the above role transition, E1 continued to track the role transition in the following snapshot, so he clicked on the informal role (\autoref{fig:case21}). All the transition flows connected to the roles in the following snapshot were highlighted. He observed one transition flow with the most significant shifts, so he just clicked on the two informal role glyphs (\autoref{fig:case21}) and the transition flow for detailed examination. After exploring the event interaction view, E1 found the same player as above. On the contrary, this player shifted his position from the center to the margin area, and all his/her network metric values dropped. Echoing the above hypothesis, E1 validated the conclusion by saying that ``\textit{when a player's all graph metrics decrease or increase, he/she goes from the center/margin to the margin/center region.''}

\par \textbf{Interaction events behind such role transitions.} To further investigate the reasons behind such role transitions, E1 compared the event distribution of this player and observed that the decline in carbon and the increase in battle jointly lead to such transitions. Respectively, what E1 found told that too many battles (competition matters) might damage players' social role while sufficient carbons (cooperation matters) may benefit their social positions.

\subsection{User Study}
\par We also conducted a user study to further assess the potency of \textit{RoleSeer}. Specifically, we evaluate our system in two aspects: \textit{workflow effectiveness} and \textit{convenience of visual design}.

\par \textbf{Participants}. We recruited $12$ participants to perform a user study ($4$ females,  $8$ males, $age_{mean}=24$,  $age_{sd}=5.196$). They are $11$ students and a teacher from the majors of computer science and game entertainment in a local university through word-of-mouth. We chose the participants with MMORPG gameplay experiences for which they could provide us with more comprehensible insights and help verify the system usability. After a brief introduction, all participants could easily understand the background and motivation. The study was conducted face-to-face, where the participants would take tutorials, experience the system as they wished, and fill out the questionnaire. After completing the user study, they would receive a \$10 voucher for an online e-commerce platform.

\par \textbf{Tasks.} Participants were asked to perform the following four tasks sequentially, which guided them to go through \textit{RoleSeer}'s pipeline. \textbf{Task One} was to examine the overall evolution dynamic of the gameplay social network in order to narrow down timestamps of interest for exploration. \textbf{Task Two} was to understand the characteristics of the informal roles and trace their evolutions across timestamps, i.e., identify a continuous role change, in the role transition overview. \textbf{Task Three} was to determine the differences between two informal roles and evaluate whether the role transition projection facilitates comparison at a group level. \textbf{Task Four} was to describe an individual player in his/her interaction behaviors and compare two players in their gameplay characteristics. Finally, after finishing all tasks, participants must complete a questionnaire with 7-point Likert scale questions (\autoref{fig:user_study}).

\par \textbf{Procedure.} The whole process for one participant lasted about $40$ minutes. Before we introduced our system, all participants were acknowledged that their reactions and feedback would be anonymously collected in the questionnaire. We then conducted a tutorial session for about $10$ minutes, in which we introduced the workflow of our system, explained all the interactions in each view, and told participants the meanings of visual encodings. Then, we allowed them to pre-operate the system and consult us for further questions for another $10$ minutes. When the participants got ready, they were required to conduct the tasks mentioned above with our system. Participants were also asked to think aloud their ideas when performing all the tasks. After finishing all the tasks, participants must complete a questionnaire with 7-point Likert scale questions. The measurements in the questionnaire included $11$ items about our system's workflow effectiveness and five items about the convenience of visual design~\cite{sun2020dfseer, xia2019peerlens}. Other comments not involved in the questionnaire were also recorded for reference.

\par \textbf{Results.} The questionnaire results are presented in \autoref{fig:user_study}. In general, our system was highly rated. Specifically, in terms of workflow effectiveness, the participants commented that \textit{RoleSeer} shows sufficient information at a time level (Q1), but the Tran is too low to be witnessed (Q2). P1 (female, age: $21$) commented that ``\textit{for the Nodes, Edges, and Tran, line chart may be a better visualization.}'' The participants felt \textit{RoleSeer} could clearly show the characteristics of the informal roles (Q3), facilitate comparison (Q4), and can help trace the long-term evolution of informal roles by tracking the visualization of the Sankey-like chart (Q5). The participants all praised the group-level comparison since it could fill the gap between the role-level and the individual-level analysis (Q6, Q8). However, some participants thought that the design of the metric statistics view did not help much in the workflow, ``\textit{it only presents some metric information for comparison}'' (P3, male, age: $22$) (Q7). The participants also found it interesting to explore the player's sequential events and community, where they could find patterns related to real-life (Q9, Q10, Q11). Regarding the convenience of visual design, the participants agreed that while most of the benefits of \textit{RoleSeer} visualizations can be attained with a specific learning curve, they could develop their paths to explore with the system (Q13). They all agreed that \textit{RoleSeer} is quite helpful in finding the roles of interest (Q12), inspecting informal roles (Q15), tracing and understanding role changes (Q16). The participants also thought that the workflow from the overall to the fine-grained provided by the visualizations was intuitive and facilitated their analysis (Q14).

\subsection{Expert Interview}
\par We conducted a semi-structured interview with the experts to evaluate \textit{RoleSeer} and to check whether our approach helps them inspect and understand the informal roles in the MMORPG context.

\par \textbf{System Performance.} We asked the experts to evaluate the results of our back-end engine, e.g., the characteristics of the identified informal roles since we did not have labeled ground truth. E3 reported that ``\textit{the abstract graph metrics are insufficient to provide insights to concrete game design, but we can see the differences by the in-game metrics,}`` echoed by E4. Through the clustering analysis, they confirmed that \textit{``the metric statistic effectively help us dig out the concrete reasons that lead to role changes in a comparative manner.``} In a word, \textit{RoleSeer} provides game designers with more concrete information to explain informal roles generated by the back-end model. Furthermore, our manner demonstrates the capability of revealing underlying interaction patterns that contribute to role changes are meaningful compared to what the experts can identify manually, ``\textit{the coverage of behavioral patterns can facilitate sense-making of role changes and game designs}''. 

\par \textbf{Visual Design and Interaction.} All the experts appreciated the ability of \textit{RoleSeer} to support the interactive exploration of the informal roles. Conventionally, the game team needs to manually distinguish different levels of players and visualize the social network at different time steps for comparison. ``\textit{We often spend days on the entire process since we need to extract different players,}'' said E1. \textit{RoleSeer} can display multiple time snapshots in real-time, making it easier for the experts to interact with the data instead of going through the raw dataset, and largely shortening the analysis time. \textit{RoleSeer} further streamlines the process as the experts no longer need to randomly select players and specify the sudden changes in the social network by hand. Instead, all relevant interaction patterns are automatically retrieved from a collection of players who experience role changes. E2 was satisfied with the system's ability to extend the analysis scale and achieve sufficient patterns behind different role changes, ``\textit{it helps me learn the significant behavioral patterns that play an important role in the role change and quickly identify similar players to confirm my assumptions}''. With \textit{RoleSeer}, the team was able to identify patterns for typical role changes. 

\par \textbf{Takeaway.} The experts are surer of the importance of a good social network for games and players. Through \textit{RoleSeer}, they find that players holding the marginal informal roles are usually limited to specific behavioral patterns such as a significant percentage of \textit{battle} events, and only interacting with a small number of players is not conducive to the development of a good social network. ``\textit{A good social network should be a network with a low proportion of players holding the marginal informal roles,}'' said E2. The diversity of interactions can ``\textit{reduce the likelihood of loss due to certain broken social relationships and improve the efficiency of integrating into roles, e.g., with sufficient \textit{carbon} events,}'' said E4. In addition, the experts further identify that encouraging role-to-role changes is a sign of a good social network, i.e., the ``openness'' of a social network can accommodate players' role changes, such as newcomers can gradually hold specific ``core'' informal roles.

\par \textbf{Generalizability and Scalability.} When discussing which component(s) of \textit{RoleSeer} can be directly generalized to other application scenarios and which need(s) customization to explore the potentials of our system, two insights were obtained: \textit{(1) Visual Design.} The experts commented that our designs, especially the role transition view and the individual interaction view are pretty generic and can fit other similar real-world scenarios. \textit{(2) Data.} \textit{RoleSeer} covers the typical data and logs that the MMORPGs can collect and do not rely on particular data types, which can be easily modified to attach to other existing MMORPGs' social network analyses. In terms of scalability, we preprocess the social network to multiple snapshots and conduct the exploratory analysis at multiple levels. Furthermore, we do not directly render the nodes and edges of the network; thus, we resolve the scalability issue to some extent.

\section{Discussion}
\par In this section, we discuss the high-level synthetics, the design implications, and the limitations of our approach.

\par \textbf{Contributions Over Previous Work.} Previous studies suggest that a social network with few marginal players and low churn risk should be stable. Our work confirms that the social establishment of players has a clear correlation with their subsequent retention, i.e., a rather persistent player social network tends to have a low percentage of marginal nodes, and most members in the network should have more than one social relationship. Our work also sheds light on the possible reasons as to why the ``decentralized'' groups gradually developed from small groups perform better in terms of stability and openness. 1) It is easy to establish multilateral relations within small groups~\cite{li2017visual}. As a result, the proportion of marginal players is small, and the overall stability is improved (\autoref{fig:community}). 2) The social network of a small group is inclined to accommodate newcomers, mainly reflected in two aspects: a higher proportion of newcomers in the entire network; and the possibility for newcomers to slowly change or develop their social roles into more central positions. 3) Compared with the core group, the threshold for newcomers to be integrated into a small group is lower~\cite{li2017visual,lu2019visual}, e.g., through a mentor-apprentice relationship or fighting against game dungeons with existing members. Furthermore, because the network is developed from multiple small groups, newcomers are more likely to hold relatively central positions as the network develops~\cite{li2017visual}. 4) There exist certain social roles~\cite{ang2010social,canossa2019influencers,williams2014structural, gladwell2006tipping} in small group social networks to enhance the gaming experience. A typical example is a social role ``connector''~\cite{gladwell2006tipping} defined as the shared friend of multiple players who plays the role of passing messages to a different community. Our work discovers a special type of ``connector'', i.e., ``bridge-builder'', who also connects players from different communities but better facilitates relationship establishment among second-order friends (i.e., ``alters''). The friends of ``bridge-builder'' are likely to play together and interact with each other directly, as shown in ~\autoref{fig:case1}. Thereby, a significant contribution of \textit{RoleSeer} over the theory of ``connector'' is that we further identify the role of ``bridge-builder'', who is critical in establishing friendship.

\par \textbf{Implications for Promoting Socialization in MMORPGs.} This work gives several design implications for socialization promotion in the virtual world. \textit{1) Promote social establishment.} This study found that even if there is only one social relationship with limited interactive events, e.g., playing in a dungeon together, a player's retention rate increases by $10$ -- $20$\% (\autoref{fig:retention}). This suggests the need to stimulate players to socialize, break the ice, and establish the first pair of social relationships, possibly through ``participating in dungeons or missions with friends'' or ``guiding players to interact with their friends''. \textit{2) Promote the growth of the number of relationships.} In \autoref{fig:retention}, we found that if players can have $5$ -- $7$ effective social relationships early in their game experiences, their retention rate increases to more than $95$\%. Thus, it is beneficial to increase the number of social relationships a player has through methods such as by adding an information push like ``you have not participated in the dungeon with your friend for a long time''. Also, it could be helpful to encourage players to maintain social relationships actively, e.g., by providing multiple pairs of mentor-apprentice relationships. \textit{3) Promote the integration of small groups.} A ``decentralized'' social network established by integrating small groups is more stable and accessible to newcomers than a ``centralized'' social network guided by a core group~\cite{li2017visual}. This could be achieved by approaches including ``breaking up the fixed team'' or ``introducing friends to other friends for rewards''.

\par \textbf{Generalization to Beyond MMORPGs.} \textit{RoleSeer} could be applied to other scenarios like \textit{user churn prediction}. User churn denotes the phenomenon where users stop using an online service, and early detection/prediction of churning is crucial for the online system's maintenance and development. When users begin to leave, their daily activeness on the platform might gradually or sharply drop and such status change may lead to informal role change~\cite{yang2018know}. \textit{RoleSeer} could help identify the likelihood of user churn before it happens by visualizing how the status of the social role changes over time. Following the properly designed visual tips, the experts could easily track how different social roles evolve, inspect their reasons, further correlate the individual behaviors with user churn and explore how users' ego community influences their retention. Such retrospective analysis can assist experts in discovering user churn patterns and diagnosing online service designs.

\par \textbf{Limitation.} This work has several limitations. First, we only worked with a relatively small team of experts (E1 -- E4) for the expert interview evaluation, and we recruited a limited number of participants to evaluate the workflow effectiveness and the efficiency of visual designs in a qualitative manner. Second, we adopt only ten key metrics to characterize the detected informal role cluster, and more metrics such as centrality and within module degree mentioned by existing work~\cite{bader2003automated,hirsch2005index,wu2008w,li2018embeddingvis} could be explored in the future. Third, this work only focuses on one-step role change, i.e., inspecting the players' interactive behavior patterns from one informal role to another without considering a sequence of role changes down a path in the player social network.

\section{Conclusion and Future Work}
\par This study presents a framework that allows us to interactively inspect the potential informal roles and explore the underlying patterns that lead to the interconversion, transitions of roles in MMORPGs. Specifically, we first propose a novel methodology based on structural-similarity-preserved network embedding and alignment to computationally model each player. We then develop four major visualizations that enable the output of the back-end model to be quickly inspected at the overview, role cluster, and individual levels. Empirical case studies and experts' feedback verify the efficacy of our approach. For future work, we will consider sequences of role changes that may depict a collection of players' entire trajectories in the game instead of only focusing on role changes between two adjacent time snapshots. Also, we will perform a systematic evaluation involving more experts.

\begin{acks}
We are grateful for the valuable feedback and comments provided by the anonymous reviewers. This work is partially supported by the research start-up fund of ShanghaiTech University and the Research Grants Council of the Hong Kong Special Administrative Region, China under Grant No. T44-707/16-N.
\end{acks}

\balance
\bibliographystyle{ACM-Reference-Format}
\bibliography{sample-base}


\end{document}